# Metalens-transparent ultrasound transducer module for extended depth photoacoustic microscopy

Krishnendu Samanta[1], Myunghoo Lee[2], Shubham Mirg[1,3], Zhihao Zhou[2], Saniya Patil[1], Johannes E. Fröch[2], Arka Majumdar[2,4], and Sri-Rajasekhar Kothapalli[1,3,5]*

[1]Department of Biomedical Engineering, Penn State University, University Park, PA 16802, United States

[2]Department of Electrical Engineering, University of Washington; Seattle, WA 98189, United States

[3]Center for Neural Engineering, Penn State University; University Park, PA 16802, United States.

[4]Department of Physics, University of Washington, Seattle, Washington 98195, United States

[5]Penn State Cancer Institute, Penn State University; Hershey, PA 17033, United States.

*Corresponding author. Email: srkothapalli@psu.edu

**Abstract**

Optical-resolution photoacoustic microscopy (ORPAM) provides label-free optical-absorption contrast at millimeter-scale tissue depths with micrometer-scale lateral resolution. However, its volumetric coverage is limited by the short depth of focus of conventional optical lenses, whereas system miniaturization is constrained by bulky optics and ultrasound transducers. Here, we present an extended-depth ORPAM system that combines an ultra-thin metalens for optical excitation with a planar transparent ultrasound transducer (TUT) for photoacoustic detection. Fabricated from piezoelectric lithium niobate, the chip-scale TUT provides ~80% optical transparency and enables coaxial optical excitation with simultaneous acoustic detection at 14.4 MHz center frequency. This architecture eliminates the need for bulky acousto-optic combiners and large-volume acoustic coupling arrangements between the transducer and imaging target. We evaluate three ultra-thin metalens designs as hyperbolic, quadratic, and extended depth-of-focus (EDOF), which provide effective axial ranges of 0.3, 0.8, and 1.3 mm, respectively, with lateral resolutions of 1.1 µm for hyperbolic, 1.2 µm quadratic lenses, and 1.5 µm for EDOF across the full axial range. Phantom experiments demonstrate improved visualization of multilayered, inclined, volumetrically distributed targets with the EDOF design compared with the two other metalenses. Label-free imaging in awake, head-fixed mice further demonstrates that EDOF provides extended-depth visualization of subcortical microvasculature without being subjected to anesthesia-induced confounds. This chip-scale metalens-TUT architecture provides a foundation for developing compact photoacoustic microscopy systems.

## INTRODUCTION

Optical-resolution photoacoustic microscopy (ORPAM) combines diffraction-limited optical excitation with ultrasound detection to visualize optical absorption at micrometer-scale lateral resolution from millimeter-scale tissue depths (*1–6*). Absorption of nanosecond laser pulses by endogenous chromophores such as hemoglobin, melanin, lipids, nucleic acids, and collagen generate broadband photoacoustic waves that are detected beyond the ballistic optical regime. Consequently, label-free ORPAM has enabled diverse pre-clinical and clinical applications, including histopathology of tissue specimens using ultraviolet light absorption of nucleic acids (*7*); *in vivo* imaging of oxy- and deoxy-hemoglobin dynamics down to single red blood cell resolution (*8*, *9*); handheld devices for imaging microvasculature of human skin and oral cavity; studies of tumor angiogenesis in rodent models (*10*, *11*), and mapping of cerebral microvasculature and oxygenation (*9*, *12*, *13*).

Despite these advances, conventional ORPAM remains constrained by two coupled engineering limitations. First, the high numerical-aperture (NA) optical lenses generate a tight Gaussian focus (*7*, *9*, *12*, *14*) required for high lateral resolution in the range of 1-10 µm. However, this leads to a short Rayleigh range and restricted depth-of-focus (DOF) of ~200 µm. Targets outside this narrow focal region receive reduced optical fluence and are imaged with degraded lateral resolution and signal intensity. This limitation is particularly consequential for volumetric specimens and curved biological surfaces. Mechanical translation of the optical focus along the axial direction can restore lateral resolution at depth but increases acquisition time and susceptibility to motion artifacts. Conversely, low-NA configurations extend the DOF only at the expense of lateral resolution and image fidelity. Second, conventional ORPAM systems generally use optically opaque focused ultrasound transducers. Coaxial alignment of optical excitation and acoustic detection paths therefore requires a coupling arrangement that limits the compactness and spatial flexibility of the complete ORPAM system.

Several optical beam-shaping approaches have been investigated to extend the DOF of ORPAM (*15*, *16*). These include axicons (*17*), Bessel beam illumination (*18*, *19*), spatial light modulators (*20*), dynamically tunable focusing (*21*, *22*), and synthetic aperture or computational refocusing methods (*23*, *24*). Although these approaches can increase the effective imaging range, they are often limited by sidelobe interference, reduced imaging speed, or rely on complex post-processing. More recently, a customized diffractive optical element is combined with a conventional refractive objective lens to synthesize a needle-shaped excitation beam for volumetric imaging of tissue specimens and in vivo mouse cerebral vasculature (*25*). Although this approach substantially extends the focal region while maintaining a narrow central lobe, it requires precise alignment of bulky optical components and transducer detectors.

Metasurfaces (*26*–*29*) provide an attractive alternative for controlling optical phase within an ultrathin array of subwavelength scatterers. Their spatially varying phase response enables metalenses to generate tailored optical-field distributions and beam profiles that are difficult to achieve using conventional refractive optics. Several recent studies have explored the application of meta-optics to the ORPAM (*30*–*32*) imaging. An ultraviolet metalens based on Nijboer–Zernike wavefront shaping extends the DOF to approximately 220 µm while maintaining micrometer-scale lateral resolution (*30*). An axially multifocal metalens distributed optical excitation across multiple focal planes for volumetric imaging of neuromelanin in live brain organoids (*31*), and another meta-optic ORPAM implementation incorporated a phase-adapted lens to increase the focal range (*32*). While these studies establish the utility of meta-optics for ORPAM, their applications were primarily limited to phantom or cell studies using transmission geometry, in which the sample was placed between the metalens and ultrasound transducer. However, the reflection geometry is required for in vivo applications. In addition, these studies only address the challenges associated with the optical excitation path and continue to use a conventional ultrasound detector for photoacoustic detection. Because the conventional ultrasound transducers are opaque and bulky, the optical excitation path and acoustic detection path require precise alignment via acousto-optic combiners comprising prisms or complex off-axis configurations (*2*). This necessitates a large acoustic coupling arrangement, usually a water-filled tank placed on imaging subject, which complicates longitudinal imaging scans, increases mechanical footprint, and severely limits spatial freedom. These limiting factors prevent widespread use of ORPAM for imaging awake subjects. For example, in standard ORPAM setup, mice are typically imaged under anesthesia for lengthy time periods, which profoundly alters baseline vascular hemodynamics and neurophysiology (*33*). Translating ORPAM from its conventional bulky setup to a miniature system is therefore essential to eliminate both the physical burden of acoustic coupling and the physiological confounds.

Transparent ultrasound transducer (TUT)(*34*) offers a complementary solution to the acoustic-detection limitations of conventional ORPAM systems. Because light can be delivered through the active transducer aperture, a TUT enables coaxial optical excitation and photoacoustic detection without optical shadowing or the need for an acousto-optic combiner. In our previous work, we developed a piezoelectric lithium-niobate based TUT and demonstrated both ORPAM and optical imaging of cerebral vasculature in awake rodents (*35*). In this configuration, TUT also served as chronic cranial window, while conventional refractive optical lens focused excitation light through the TUT to achieve diffraction-limited imaging of subcortical brain vasculature. The optical focus was raster-scanned through the TUT using a high-speed galvanometric scanner to generate volumetric images of cerebral microvasculature. More recently, several TUT-based ORPAM systems have been developed for preclinical and clinical applications (*36*, *37*), including focused TUTs incorporating acoustic lenses to improve detection sensitivity and bandwidth (*38*, *39*). However, because these systems continue to rely on conventional refractive focusing optics, their volumetric imaging coverage remains limited by the short optical DOF.

In this work, we combine an ultrathin, phase-engineered metalens with a planar lithium-niobate TUT to establish a chip-scale optical excitation-acoustic detection module that enables compact extended-depth ORPAM of imaging awake subjects. To systematically evaluate how metalens phase engineering affects volumetric ORPAM performance, we fabricated and compared three different metalenses: hyperbolic, quadratic-phase, and extended-depth-of-focus (EDOF). The hyperbolic and quadratic-phase designs use analytically prescribed phase profiles, whereas the EDOF metalens uses an optimized aperiodic radial phase to generate an extended axial focus through the TUT into the subcortical brain regions. Specifically, in the EDOF metalens, we exploit the intrinsic chromatic dispersion of the meta-optic, for which the product of focal length and wavelength stays nearly constant (40). We fix the focal plane at the nominal focal distance and optimize the phase profile to maximize the minimum on-axis intensity that a band of design wavelengths delivers to this single plane. Because a change in wavelength is equivalent to a shift of the focal plane, at the single 532 nm excitation wavelength this optimized profile forms a continuous needle-shaped focus with near-uniform intensity over the targeted axial range in the subcortical brain region. The metalens delivers the excitation beam through the optically clear 1 mm thick TUT, which facilitates raster scanning of the beam and simultaneous detection of photoacoustic signals through the same aperture. The TUT is positioned directly above an optically and acoustically transparent cranial window using only a thin acoustic-coupling layer, thereby avoiding the large coupling volume required by conventional ORPAM configurations. Experiments in multilayered phantoms demonstrate that the EDOF metalens provides 1.5-µm lateral focal width over a measured optical DOF of 1.3 mm, compared with the shorter focal ranges of the hyperbolic and quadratic metalens designs. Imaging in awake, head-fixed mice further demonstrates extended-depth visualization of cerebral microvasculature. Together, the metalens-TUT platform establishes a planar, coaxial architecture for extended-depth ORPAM and provides a foundation for development of miniatured photoacoustic imaging probes for basic and clinical applications.

## RESULTS:

### *Concept of the technique*

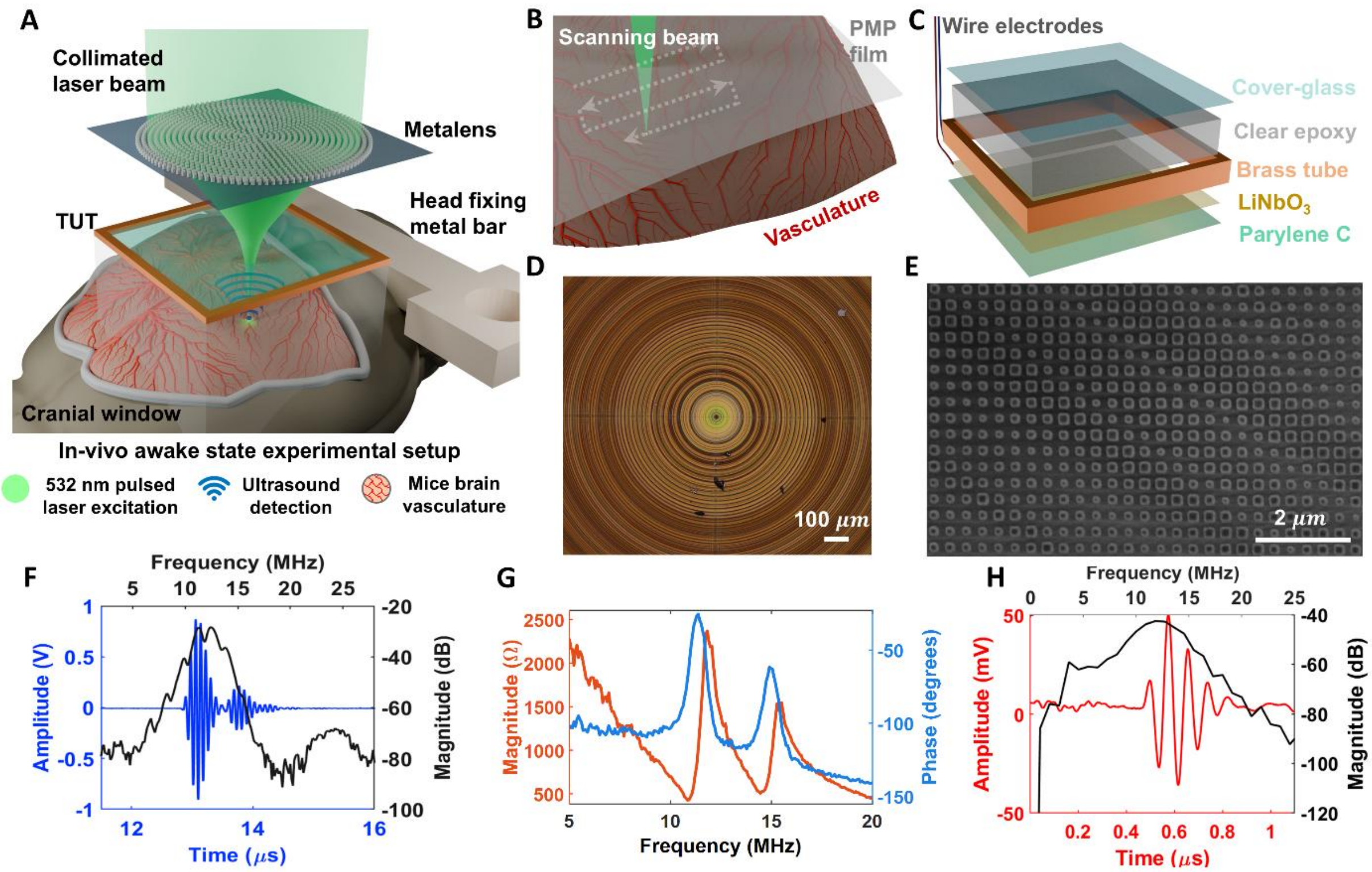

**Fig. 1. Design and characterization of the metalens-transparent ultrasound transducer (TUT) based ORPAM setup**: (**A**) Schematic of the experimental configuration showing metalens-based optical excitation, and coaxial photoacoustic detection through the TUT; (**B**) enlarged schematic view of the imaging region showing the focused excitation beam raster-scanned across the cranial PMP film window; (**C**) cross sectional view of the TUT showing its constituent layers; (**D**) brightfield optical image of the extended-depth-of-field (EDOF) metalens; (**E**) Scanning electron micrograph (SEM) of the EDOF metalens structures; (**F**) measured pulse-echo response of 3x3 mm$^2$ TUT and corresponding frequency spectrum; (**G**) measured electrical impedance magnitude and phase of the TUT as a function of frequency; (**H**) representative photoacoustic waveform acquired from a planar absorbing target (black tape) and its corresponding frequency spectrum. ORPAM: optical resolution photoacoustic microscopy, PMP: polymethylpentene.

We have developed an extended-depth ORPAM architecture that co-axially combines an ultrathin (300 µm thick) metalens for optical-excitation with a planar TUT (1 mm thick) for photoacoustic detection. The schematic of the system configuration is illustrated in Fig. 1A. As TUT allows the excitation beam to pass through its active aperture, the metalens-TUT arrangement places the optical and acoustic components along the same axis and removes the need for the bulky acousto-optic combiners commonly used to co-align separate optical and acoustic paths. A magnified view of the imaging geometry is shown in Fig. 1B, and the optical path through the component stack is detailed in supplementary S1.

### *TUT as mini acoustic detection module*

In the proposed technique, a chip-scale TUT is employed as the mini acoustic detection module. The TUT is fabricated using an optically transparent 250-µm thick lithium niobate ($LiNbO_3$) piezoelectric layer coated with transparent indium tin oxide (ITO) electrodes, a transparent backing layer, and a Parylene-C matching plus insulating layer. A cross-sectional view of the TUT device with multiple layers is shown in Fig. 1C, and detailed fabrication process is discussed in the materials and methods section and in supplementary S4. TUT exhibits broadband optical transparency (*34*) across the visible spectrum, with a transmittance of approximately 80%, measured using a power meter (S120C, Thorlabs, NJ, USA) at the 532 nm excitation wavelength.

TUT chips of size 3 mm x 3 mm x 1 mm are employed for the phantom studies whereas TUTs of size 6 mm x 6 mm x 1 mm are used for the in-vivo mouse brain experiments. The TUT is highly sensitive for detection at a close proximity (within 3 mm) to the sample and requires a negligible amount of acoustic coupling medium. Moreover, it efficiently detects photoacoustic signals without any bulky acoustic lens, which makes it suitable for light scanning through the clear aperture without mechanical translation of the imaging head. Details of the focused excitation beam transmitting through different layers of the TUT is shown in supplementary S1.

Following the assembly and addition of Parylene-C layer, we first measured electrical impedance of TUTs using a network analyzer (E5100A, Agilent, CA, USA). The impedance-phase response of the TUT shown in Fig. 1G reveals resonances at 11.6 MHz and 15.4 MHz, while corresponding antiresonances at 11.9 MHz and 15.7 MHz. The nominal thickness-mode center frequency of the bare 250-μm $LiNbO_3$ element was calculated to be approximately 14.4 MHz. Pulse-echo characterization of a 3 mm x 3 mm TUT connected through an impedance-matching circuit provides a peak-to-peak echo amplitude of ~ 1.7 V, shown by blue plot in Fig. 1F. The black trace in Fig. 1F shows corresponding frequency spectrum, with a -6 dB fractional bandwidth of approximately 29%. A representative photoacoustic waveform acquired from a planar absorbing target (black tape) and its corresponding frequency spectrum are shown in Fig. 1H. Together, these measurements establish that the transparent detector provides the optical access and acoustic bandwidth required for the ORPAM experiments described below.

***Metalens as mini optical excitation module***

The planar meta-optics is employed as the optical excitation and beam-shaping module with the smallest possible footprint in the proposed system. This approach fundamentally addresses the constraints imposed by bulky refractive components, namely their stringent optical alignment tolerances and inherent volumetric footprint. To demonstrate different types of functionalities and specifically highlight the capabilities of the EDOF metalens, we have designed three different phase-profiles - hyperbolic, quadratic, and EDOF for this study. The hyperbolic metalens designed for 3 mm diameter and the other metalenses designed for 6 mm diameter demonstrate the influence of depth of focus modulation along with the intrinsic transverse spatial resolution governed by the NA and wavelength (λ) of excitation light. The first design is based on a hyperboloid phase profile, and is engineered to achieve an ideal, diffraction-limited focusing profile, characteristic of a perfectly corrected objective lens. Its phase profile is derived from the exact hyperbolic path difference across the lens aperture given by:

$$\varphi(x, y, \lambda) = -\frac{2\pi\mu}{\lambda}\left\{\sqrt{(x^2 + y^2 + f^2)} - f\right\} \qquad (1)$$

where x, y represent the spatial coordinates, λ is the free-space wavelength of light, f denotes the focal length of the lens and μ corresponds to the refractive index of the medium. This design provides a sharply defined focal spot with a shallow DOF where the intensity profile exhibits an asymptotically decaying profile away from the focal plane. This tight axial confinement is important for maximizing the peak photoacoustic signal intensity and achieving the highest possible axial resolution that appears at the cost of shallow DOF. The second design is based on a quadratic phase profile, which approximates the focal properties based on the paraxial wave equation. Its

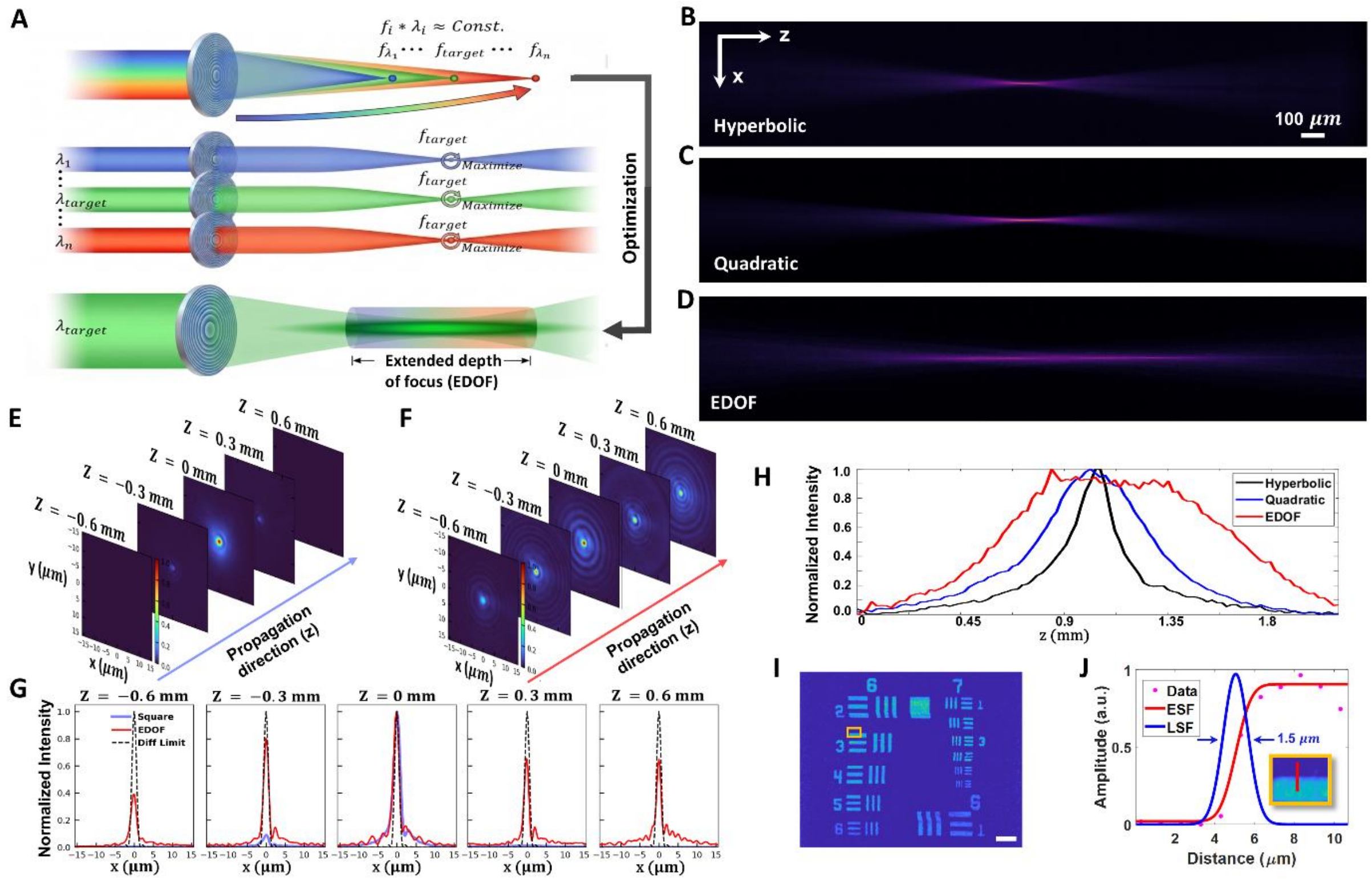

**Fig. 2. Optical and photoacoustic characterization of the three metalens designs:** (**A**) Schematic illustrating generation of an extended depth-of-focus (EDOF) excitation beam using the phase-engineered metalens; (**B** to **D**) measured axial intensity distributions (x–z) for the (**B**) hyperbolic, (**C**) quadratic, and (**D**) EDOF metalenses; (**E**, **F**) lateral (x–y) intensity distributions measured at selected axial positions along the propagation (z) direction for the (**E**) quadratic-phase and (**F**) EDOF metalenses; (**G**) lateral intensity profiles at different axial positions demonstrating the extended, near-uniform focal confinement of the EDOF metalens; (**H**) comparison of the normalized axial intensity profiles, yielding DOFs of approximately 0.3, 0.8 and 1.3 mm for the hyperbolic, quadratic, and EDOF metalenses, respectively; (**I**) ORPAM image of a USAF resolution target acquired using the EDOF metalens (scale bar, 35 μm); (**J**) edge-spread function (ESF) measured across a sharp edge of the USAF target and the corresponding line-spread function (LSF) used to estimate lateral resolution.

phase profile is represented by a quadratic function of the lateral spatial coordinates:

$$\varphi(x, y, \lambda) = -\frac{\pi\mu}{\lambda f}(x^2 + y^2) \qquad (2)$$

The phase profile of the quadratic metalens provides a longer DOF relative to the hyperbolic metalens, albeit with slightly relaxed axial precision. In this study, we use this design as a baseline comparison for the inherent trade-offs between axial confinement and signal robustness across relatively longer depth ranges.

The third design is an extended depth of focus (EDOF) metalens. Unlike the previous two metalens designs, the EDOF metalens utilizes an optimized aperiodic phase modulation across its aperture to deliberately extend the axial focal range while preserving the lateral resolution. Our design method leverages the intrinsic chromatic dispersion of meta-optics, where the product of the focal length and wavelength remain nearly constant (*40*). Because this product is conserved, a synthetic band of design wavelength is mapped onto a corresponding set of axial focal planes that together tile the targeted depth of focus; the near and far limits of the needle-shaped focus are fixed by the two end member products $f_1\lambda_1$ and $f_2\lambda_2$ so that,

$$f_i\lambda_i \approx f_0\lambda_0 = const, \quad f_i = \frac{f_0\lambda_0}{\lambda_i} \qquad (3)$$

where $f_0$ and $\lambda_0$ are the nominal focal length and design wavelength. The discretized phase map $\varphi(x, y)$ is treated as a free optimization variable. We compute the on-axis intensity at a single plane fixed at the nominal focal distance, and over a band of N design wavelength spanning the target range we maximize the smallest of these intensities. This optimization drives the axial intensity toward a near-uniform, needle-shaped profile rather than a single sharp peak,

$$\varphi^* = \arg\max_{\varphi} \min_{1\leq i\leq N} I(\lambda f_i; \varphi) \qquad (4)$$

where $I(\lambda f_i; \varphi)$ is the on-axis intensity delivered to the fixed focal plane by the $i^{th}$ design wavelength for a phase profile φ. The phase map is iteratively optimized using a gradient-based approach with the Adam optimizer to maximize the worst-case focal intensity across the corresponding design wavelengths to maintain a continuous, single-wavelength focus over the extended axial range, as shown in Fig. 2A. The wrapped phase profiles for the three metalenses are shown in supplementary S2. With respect to the achievable focus depth range, the EDOF metalens outperforms the other two designs, which inherently suffer from shallow depth of focus. This is typically achieved by trading the peak intensity at the focus for a near-uniform intensity distribution across the extended axial depth. This allows simultaneous photoacoustic excitation over a larger depth range to yield a higher depth-encoded ORPAM signal. Although we trade off some spatial resolution, the EDOF approach significantly improves volumetric imaging capability with extended depth-of-focus. The resulting ORPAM signal is a depth-encoded map for 3D volume from a single-focus acquisition. The brightfield optical microscopy image of the EDOF metalens is shown in Fig. 1D and SEM micrograph of a small region is depicted in Fig. 1E.

All three metalenses are fabricated using silicon nitride (SiN) on a quartz substrate resulting in a highly compact overall thickness of approximately 300 μm. The fabrication process of metalens is explained in supplementary S5 as well as in Materials and Methods section. Figs. 2(B-D) illustrate the experimentally characterized axial intensity distributions (x-z cross-sections) of the hyperbolic, quadratic, and extended-depth-of-focus (EDOF) metalenses, respectively. Focal depth characterization is performed using optical microscopy to assess their respective depth-of-focus (DOF) performances. The comparative analysis reveals a gradually increasing axial confinement capability: $DOF_{Hyperbolic} < DOF_{Quadratic} < DOF_{EDOF}$ with redistributed intensity along the propagation (z) direction. The hyperbolic metalens in Fig. 2B optimizes the phase profile to emulate an ideal converging spherical wavefront to achieve a diffraction-limited focal spot with a strictly constrained depth of focus (DOF) whereas the steep longitudinal phase curvature induces a rapid post-focal divergence. In contrast, the metalens with quadratic-phase profile in Fig. 2C slows down the on-axis intensity decay and offers a moderate DOF extension while maintaining a high degree of lateral confinement. Fig. 2E shows the lateral (x-y) intensity cross-sectional image stacks along multiple axial planes for the quadratic metalens to highlight its axial depth range. Most significantly, the EDOF metalens in Fig. 2D leverages a non-convex phase profile to generate a needle-shaped EDOF excitation beam for ORPAM. Fig. 2F displays the intensity image stacks at different axial locations of the EDOF metalens and confirms the extended depth range from z = -0.6 mm to z = 0.6 mm. Fig. 2G shows the lateral intensity line profiles at different axial positions, measured using an optical microscopy setup shown in supplementary S6. The average lateral intensity full width half maximum (FWHM) of 1.5 μm for EDOF metalens remains nearly invariant across the entire axial extent, confirming the extended and uniform focal confinement of the needle-shaped beam. The quadratic metalens exhibits near-uniform lateral intensity FWHM of 1.2 μm whereas hyperbolic metalens offers tightly focused beam with lateral intensity FWHM of 1.1 μm. The axial intensity line profiles are plotted in Fig. 2H to highlight the comparative DOF range and the FWHM calculation estimates the range to be roughly 0.3 mm, 0.8 mm and 1.3 mm for hyperbolic, quadratic and EDOF metalenses, respectively. A flat USAF target in agar is tested for characterizing our system using EDOF metalens and corresponding ORPAM image is shown in Fig. 2I. The horizontal and vertical intensity line profiles are presented in supplementary S10. A small area-of-interest having a sharp edge in ORPAM image is analyzed, the edge spread function (ESF) and line spread function (LSF) are plotted in Fig. 2J. Inset shows the magnified sharp edge region and FWHM calculates the lateral resolution of 1.5 μm from ORPAM signal, suitable for micro-vasculature imaging.

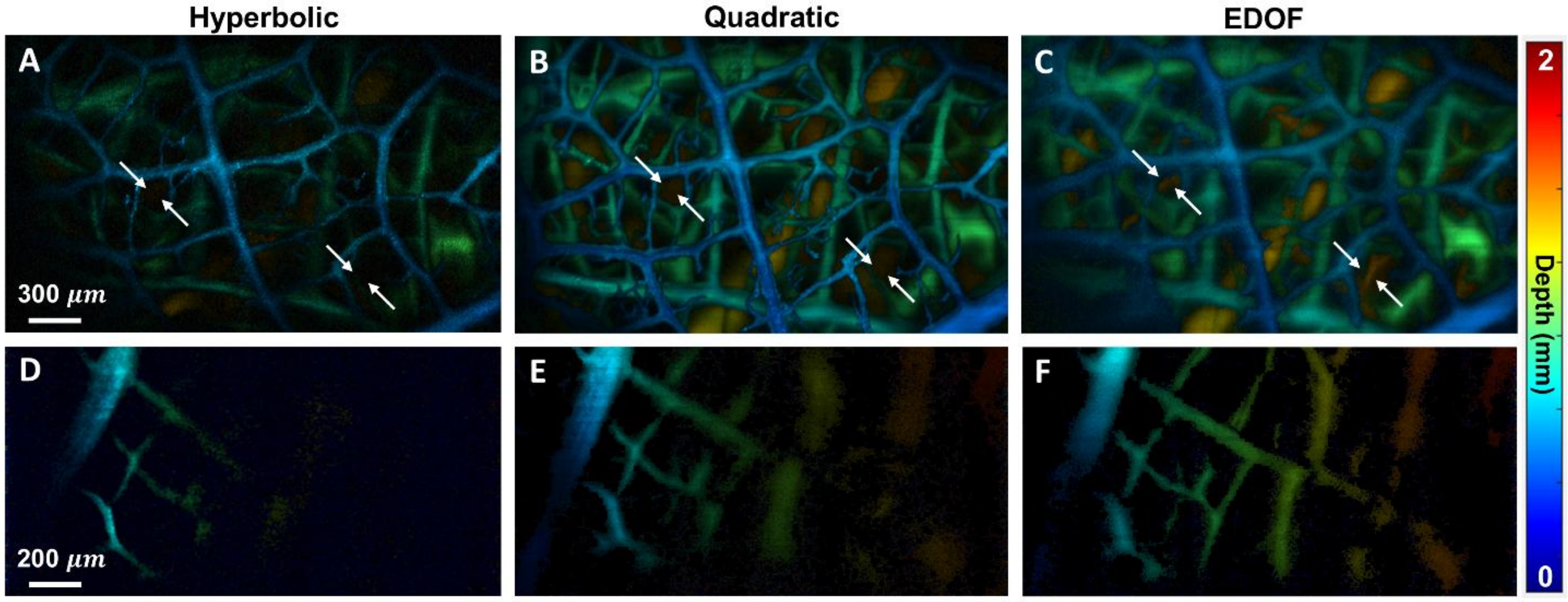

**Fig. 3. Extended-depth ORPAM imaging of multi-layered planar and single-layer inclined leaf phantoms:** (**A** to **C**) Depth-encoded ORPAM images of multi-layer planar leaf phantom acquired using (**A**) hyperbolic, (**B**) quadratic, and (**C**) EDOF metalenses, white arrows indicate structure in the third layer; (**D** to **F**) depth-encoded ORPAM images of single-layer inclined leaf phantom using (**D**) hyperbolic, (**E**) quadratic, and (**F**) EDOF metalens.

***ORPAM imaging of structurally diverse phantoms***

The proof-of-principle experimental demonstration is performed using different phantoms that are designed to emulate the complex architectures mimicking various biological environments. First, a phantom consisting of vertically stacked multiple planar leaf (mesh structures) embedded in a 1.2% agar medium is used to determine the system's ability to resolve planar interfaces across various penetration depths. Across all three metalens configurations, the technique generates photoacoustic signals throughout their respective DOFs. Importantly, it shows that the variation in DOF does not adversely affect signal uniformity or resolution. The reconstructed images show clear resolved structural features of the leaf networks, demonstrating that each metalens generates a confined optical excitation beam over its designated axial range. Fig. 3A presents the depth-encoded ORPAM image of the multi-layer leaf structure using the hyperbolic metalens. The top layer is clearly resolved and the second layer is only partially resolved, indicating a shallow depth-of-field (DOF) of 0.3 mm for this configuration. Fig. 3B shows the corresponding results obtained with the quadratic metalens, in which the top two layers are sharply resolved while the third layer appears blurred, demonstrating a moderately improved imaging depth of 0.8 mm. Fig. 3C illustrates the ORPAM result using the EDOF metalens, where all three target layers are fully resolved and confirms extended depth imaging. The structure between the arrowheads is missing in the hyperbolic metalens image, appears blurred in the quadratic metalens image, and is sharply resolved in the EDOF metalens image, demonstrating the superior depth-of-focus performance up to 1.1 mm of the EDOF design. As the multilayer leaf structure consists of inherently discrete layers, evaluating the actual depth range estimation necessitates the analysis of a continuous sample. Therefore, further experiments are performed using a phantom composed of a single inclined leaf embedded in a 1.2% agar matrix which assesses the system's capability to resolve inclined continuous targets across the axial depth range. Fig. 3D shows the experimental results of such phantom using the hyperbolic metalens where the resolved structure is extended with a shallow depth range capability of 0.4 mm. Fig. 3E represents the results using quadratic lens with improved depth range up to 0.8 mm. Similarly, the results with the EDOF metalens are shown in Fig. 3F which depicts the depth of focus is further extended up to 1.4 mm. Some complementary experimental results of such leaf phantoms using three metalenses are depicted in supplementary S8. This extended depth capability is achieved without any noticeable degradation in lateral resolution, preserving the structural fidelity of the sample.

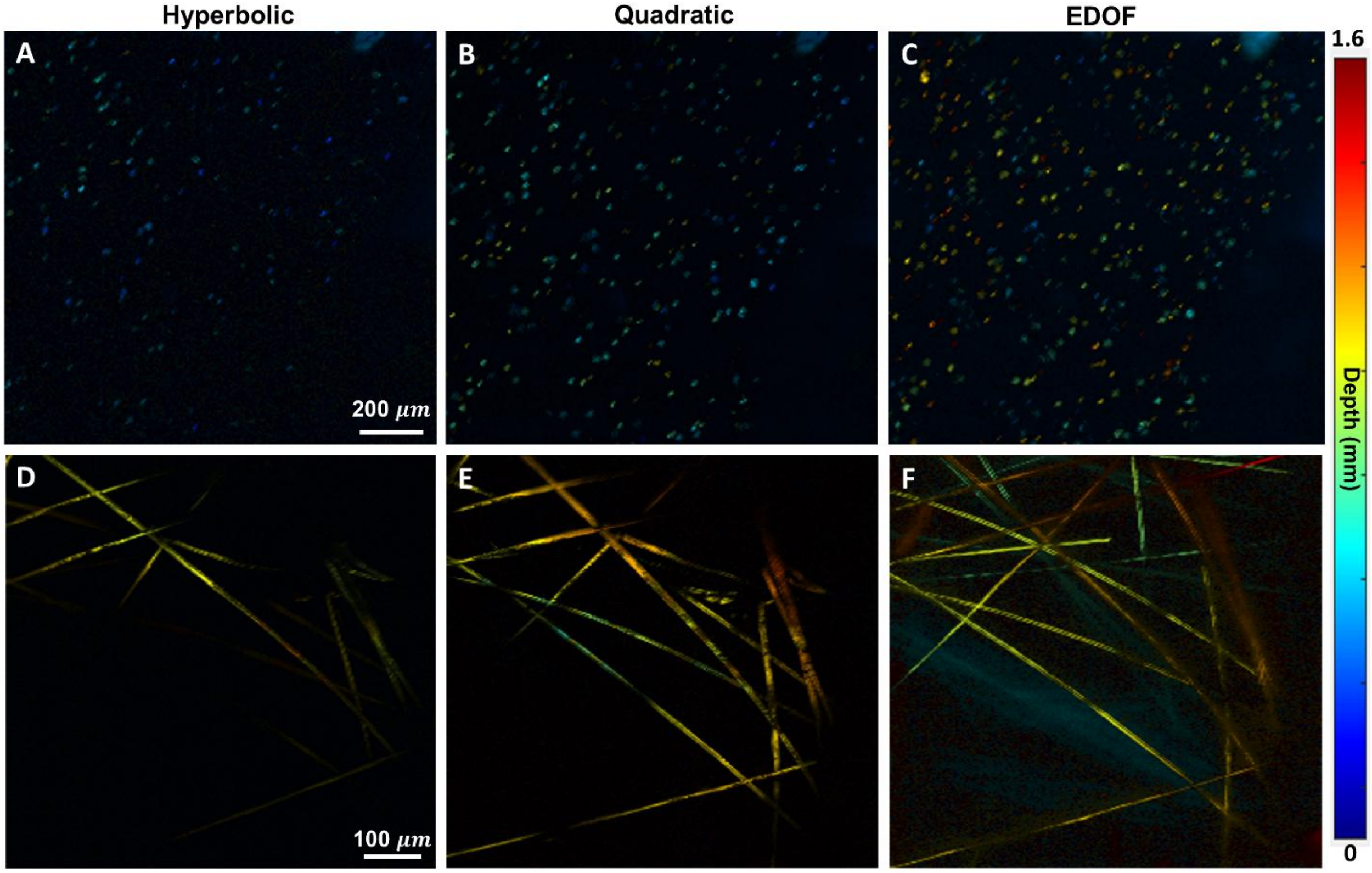

**Fig. 4. ORPAM imaging of carbon micro-particles and carbon-fibers phantoms:** (**A** to **C**) Depth-coded ORPAM images of randomly distributed carbon micro-particles (2-12 µm diameter) embedded in agar, acquired using (**A**) hyperbolic, (**B**) quadratic and (**C**) EDOF metalenses; (**D** to **F**) depth-coded ORPAM images of randomly oriented carbon fibers in agar using the (**D**) hyperbolic, (**E**) quadratic, and (**F**) EDOF metalenses.

Further, experiments are performed on randomly distributed carbon particles (Sigma-Aldrich, USA) of 2-12 µm diameter embedded in 1.2% agar. Figs. 4(A to C) present the experimental results of such phantom using the hyperbolic, quadratic, and EDOF metalenses, respectively. The number of micro-particles visualized across different depths increase progressively from the hyperbolic to the quadratic and EDOF metalenses. While the hyperbolic metalens offers minimal depth resolution of 0.3 mm, the quadratic metalens achieves relatively longer depth structures of approximately 0.8 mm. The EDOF metalens outperforms other two metalenses, providing the highest depth-resolved particles across an axial range of 1.3 mm. The results are analyzed to estimate the average lateral signal width of the carbon micro-particles, yielding values of 7.1 µm, 6.6 µm, and 8.3 µm for the three metalens designs, respectively as shown in supplementary S9. Further, we study a complex phantom consisting of carbon fibers (Asbury, MI, USA) oriented randomly within a 1.2% agar medium. This configuration presents a more rigorous challenge, simulating a volumetric environment with absorbers positioned unpredictably in three-dimensional space. The system consistently detects individual carbon fibers at varying axial positions for all three metalenses and generates photoacoustic signals across the entire supported depth range. The experimental results of this phantom for hyperbolic, quadratic and EDOF metalenses are shown in Figs. 4(D to F), respectively. The results illustrate that the increasing number of fibers appear across the depth range of 0.3 mm, 0.8 mm and 1.1 mm using hyperbolic to EDOF metalens, respectively. The results validate that selecting specific metalenses is a simple yet effective mechanism for tailoring the axial imaging window to suit different sample geometries. Collectively, these phantom experiments constitute a clear proof-of-principle demonstration that the technique enables highly flexible, depth-adaptable miniatured ORPAM imaging. By employing multiple metalenses with engineered DOFs, the system offers the unique ability to be rapidly reconfigured. It can switch between shallow, high-precision imaging and deeper volumetric interrogation without the need for mechanical z-scanning or complex optical realignment. The consistent performance observed in different phantoms (structured leaf, randomized carbon particles and fibers) suggests the potential for applying this strategy to biological samples in vivo. It is worth noting that the image acquisition rate of the current system is quite limited, primarily constrained by the slow scanning speed of <5 mm/s for the translation stages.

***Label-free ORPAM imaging of cerebral vasculature in awake mice***

Next, we demonstrate the metalens-TUT architecture for label-free in-vivo cerebral vasculature ORPAM imaging of awake mice brain. The key advantage of this miniscale ORPAM headpiece is its lightweight and compact footprint which enables stable imaging of awake murine brain without the physiological and neurovascular confounds associated with anesthesia. This feature is particularly important for photoacoustic interrogation of cerebral hemodynamics where anesthesia is known to alter vascular tone, blood oxygenation, and neurovascular coupling. A cranial window is prepared in experimental mice via craniotomy, where the skull is removed for improved acoustic transmission and the exposed cortex is sealed with 80 μm optically transparent ultrasound-friendly polymethylpentene (PMP) film. The complete surgical protocol including recovery is described in the Materials and Methods section. Supplementary S11 presents preparation of mice for the in-vivo awake-state imaging setup. For acoustic coupling and mechanical stabilization, the TUT is mounted directly above the cranial window using a 1.2% thin agar layer as an impedance-matching and coupling medium. A 6 mm × 6 mm square TUT chip is used for in-vivo studies and to avoid edge artifacts, a central 4 mm × 4 mm Region of Interest (ROI) is designated for ORPAM imaging that provides sufficient coverage for nearly one hemi-cortex. The focused laser spot modulated by the metalens is scanned across the vascular structure on the cortical surface through the TUT clear aperture. This configuration establishes miniscale coaxial optical excitation and acoustic detection through the same aperture, highly suitable for awake-state rodent studies. The depth-encoded photoacoustic image enables direct visualization of cortical microvasculature with intrinsic optical absorption contrast from hemoglobin, while simultaneously probing the effective axial confinement governed by the metalens phase profile.

The intrinsic curvature of the murine cerebral cortex presents a significant challenge for high-resolution, brain-wide ORPAM imaging. Specifically, distal vasculature located far from the superior sagittal sinus often falls outside the limited focal depth of conventional ORPAM systems and makes these regions inaccessible without mechanical scanning. To address this limitation, we have systematically evaluated the technique on the awake murine brain using three metalenses. Consistent with the axial beam characterizations, the hyperbolic metalens generates high-contrast vascular signals localized to a narrow focal plane but exhibits rapid signal decay with the axial displacement. Fig. 5A depicts the ORPAM image using the hyperbolic metalens and corresponding B-mode image across the white arrow region shown in Fig. 5E confirms shallow depth of focus of around 0.3 mm. In comparison, the quadratic-phase metalens extends the effective imaging depth while moderately preserving lateral resolution. Fig. 5B depicts the result with quadratic metalens where the distal vessels become visible and the proximal vessels appear blurred or are partially missing due to limited depth of focus. The cross-sectional B-mode image in Fig. 5F shows the improved imaging depth range around 0.8 mm. For the ORPAM image using EDOF metalens, the vessels across both proximal and distal regions appear prominently. The result with the EDOF metalens shown in Fig. 5C demonstrates nearly invariant lateral focal width across an extended axial range of 1.3 mm and corresponding B-mode image is presented in Fig. 5G. Furthermore, other awake murine brain regions near the sagittal sinus area are also imaged; the results for two mice with EDOF lens and corresponding brain-wide benchtop optical images are shown in supplementary S12 illustrating areas where the skull exhibits different curvatures. Overall, the proposed technique using the EDOF metalens facilitates high-resolution elongated depth resolved imaging micro-vessels across the curved cortical surface. Collectively, these in vivo results validate that phase-engineered metasurfaces can directly tailor the axial response of ORPAM systems under physiologically relevant conditions. The lightweight, alignment-tolerant architecture is particularly suited for awake imaging paradigms, underscoring the potential of metalens-based photoacoustic platforms for translational neuroimaging and compact biomedical diagnostics.

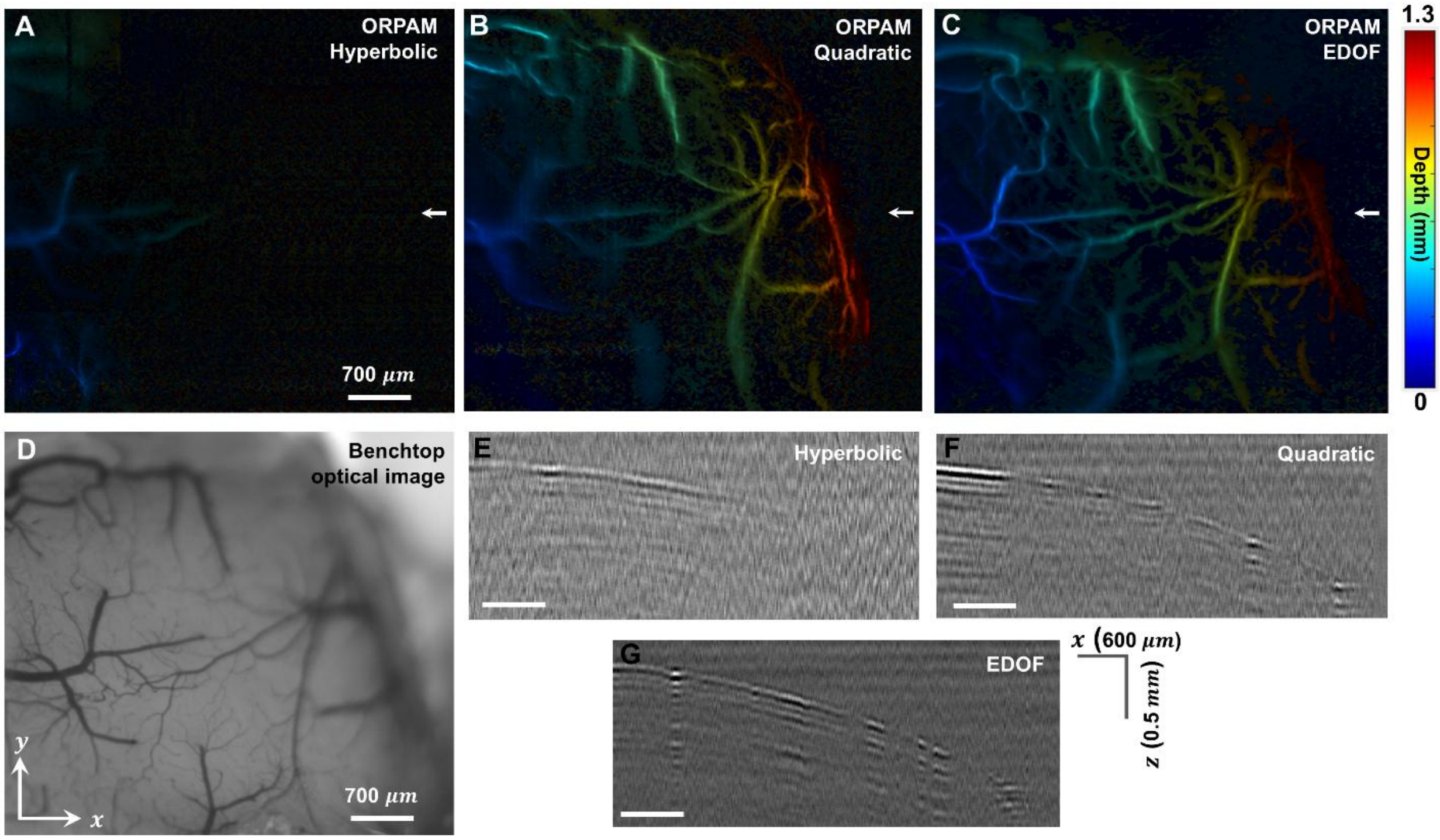

**Fig. 5. Extended-depth ORPAM imaging of cerebral microvasculature in an awake mice:** (**A** to **C**) Depth-encoded ORPAM images of brain vasculature from distal to the sagittal sinus acquired using (**A**) hyperbolic, (**B**) quadratic, and (**C**) EDOF metalenses; (**D**) optical image of the corresponding region-of-interest; (**E** to **G**) cross-sectional B-mode images along the white arrow indicated in (**A** to **C**), respectively, illustrating the axial imaging ranges of the three metalenses. Scalebar: 600 μm.

## DISCUSSION:

In this work, we have developed a compact and planar metalens-TUT architecture for extended-depth ORPAM and demonstrated its performance in phantoms and in the awake mouse brain imaging. The central feature of the approach is the combination of two complementary planar components: an ultrathin dielectric metalens that controls the axial distribution of the optical excitation field and a chip-scale transparent piezoelectric detector that enables coaxial optical transmission and photoacoustic detection. This configuration significantly reduces the optical and acoustic footprint at the imaging target while eliminating the need for a conventional acousto-optic combiner between the excitation and detection paths. As a result, this fundamentally overcomes the bulky mechanical complexity and alignment sensitivities that have long constrained conventional refractive optical-resolution photoacoustic microscopy systems.

Among the three phase profiles evaluated, the EDOF metalens phase profile generates largest optical depth of focus - a needle-shaped excitation beam through the TUT into the imaging target. This extends the measured axial range to ~1.3 mm while maintaining an average lateral resolution of 1.5 μm, effectively decoupling lateral resolution from depth of focus. The hyperbolic and quadratic-phase metalenses produce relatively shorter focal ranges of approximately 0.3 mm and 0.8 mm, respectively. This trend is reproduced across structurally diverse phantom configurations including stacked leaf meshes, randomly distributed carbon fiber networks and micro-particles (2-12 μm). Although the exact effective imaging range depends on sample geometry, the EDOF configuration consistently visualizes absorbers across a larger axial span than the other two designs. The results therefore show that phase engineering can substantially relax the conventional trade-off between lateral optical confinement and depth of focus in ORPAM over the range investigated in our study.

The capabilities of the three metalenses are further evaluated for label-free imaging of cerebral microvasculature in awake, head-fixed mice. The curved cortical surface naturally distributes vessels across different axial positions and therefore challenges the limited focal range of conventional ORPAM. The EDOF metalens provided vascular visualization over an approximately 1.3-mm axial span, compared with approximately 0.3 mm and 0.8 mm for the hyperbolic and quadratic-phase configurations, respectively. Importantly, these measurements are obtained without general anesthesia during image acquisition. The TUT is positioned directly above an acoustically compatible PMP cranial window with only a thin agar coupling layer, providing a practical configuration for ORPAM imaging in awake animals. These experiments establish the feasibility of combining phase-engineered flat optics with transparent acoustic detection for extended-depth in vivo ORPAM.

While the current study addresses the key challenge of ORPAM, that is to extend the axial imaging range using compact planar imaging head, the overall system performance and miniaturization can be further improved through the following steps. First, the current experimental system still uses external free-space beam-delivery optics and a two-axis mechanical translation stage with scanning speeds of less than 5 mm/s. Like in the state-of-the-art ORPAM system, the use of fiber optics for light delivery and integration with galvanometric, MEMS, or other miniaturized scanning approaches could substantially improve acquisition speed while retaining the planar metalens-TUT geometry (*9*, *41*). Second, extension of the optical focal range redistributes optical energy along the axial direction. The present EDOF design therefore represents a balance among focal range, lateral confinement, peak intensity, and optical efficiency rather than a lossless extension of depth of focus. Further optimization of phase profiles, focusing efficiency, sidelobe suppression, and the optical properties of the complete metalens-TUT architecture can improve this balance. Similarly, the TUT used here is designed around a nominal 14.4 MHz central frequency; choosing a relatively higher (30 MHz) center frequency transducer would help improve both spatial resolutions and sensitivity. Leveraging the scalability of TUT fabrication, future transducer designs can therefore be optimized for the intended imaging depth and spatial scale. In parallel, flexible metalenses (*42–44*) integrated directly onto the TUT surface, can facilitate a monolithic opto-acoustic module for more conformal and wearable applications, whereas the electrically tunable metalens (*45–47*) may provide dynamically tunable focal depths. Additionally, integration of this module with miniaturized endoscopic configurations (*48*, *49*) can further reduce the distal probe dimensions. Finally, extending the current single-wavelength implementation to multispectral excitation can provide functional measurements such as hemoglobin oxygen saturation in addition to vascular morphology (*8*, *41*). Collectively, these developments can transform the present architecture into a more compact versatile platform for photoacoustic microscopy, covering both pre-clinical and next-generation clinical endoscopic modalities.

## MATERIALS AND METHODS:

### *Metalens design and simulation*

The three metalenses employ square silicon nitride (SiN) nanopillars on a quartz substrate with a fixed pitch of 270 nm and a pillar height of 500 nm. A meta-atom library is built by finite-difference time domain (FDTD) simulation of a single periodic unit cell at the 532 nm operating wavelength by sweeping the pillar width from 50 nm to 210 nm. The simulated pillars provide a full 2π phase span with high transmission (supplementary S3), and each target phase is realized by selecting the pillar width that reproduces it. For the hyperbolic and quadratic metalenses, the target phases follow the analytic profiles described in Eq. 1 and Eq. 2. For the EDOF metalens, the phase profile is obtained by gradient-based optimization. The aperture phase is discretized on a radial grid of 11,111 points across the 3 mm radius and treated as a free variable initialized to a random distribution. The metasurface is modeled as a fixed structure whose imparted phase scales with the inverse of the wavelength, so that a single element focuses different wavelengths at different axial planes. The on-axis field at the nominal 15 mm focal plane is computed with the Rayleigh–Sommerfeld diffraction integral for a band of 600 design wavelengths from 509.9 nm to 556.1 nm centered at 532 nm. Because the product of focal length and wavelength is conserved, this wavelength band corresponds to a focal range of about ±0.65 mm about the nominal focus, matching the targeted 1.3 mm depth of focus. The phase profile is optimized to maximize the minimum on-axis focal intensity across this wavelength band, a worst-case criterion that produces a near-uniform axial response rather than a single sharp peak. The optimization is implemented in TensorFlow with automatic differentiation and uses the Adam optimizer with a learning rate of 0.005 for 30,000 iterations. The resulting wrapped phase is then mapped to pillar widths through the FDTD library, and the wrapped phase profiles of the three designs are shown in supplementary S2.

### *Fabrication of metalens*

The metalens fabrication process is started with the deposition of a silicon nitride (SiN) thin film onto a quartz substrate. The SiN film is deposited using plasma enhanced chemical vapor deposition (PECVD) to achieve precise thickness and uniformity. A positive electron-beam resist (ZEP-520A) is then spin-coated onto the SiN surface. Next, the desired nanostructure pattern is transferred onto the SiN film using Electron Beam Lithography (EBL) followed by the deposition of a hard mask to protect the SiN during the subsequent etching step. For the creation of a hard mask, a layer of aluminium oxide ($Al_2O_3$) is deposited by electron-beam evaporation (EVAP). A lift-off process is then performed to remove the sacrificial resist and excess metal, leaving the patterned $Al_2O_3$ mask on the SiN layer. Finally, the pattern is etched into the SiN through inductively coupled plasma (ICP) etching, forming the array of square SiN nanopillars (270 nm pitch, 500 nm height) that constitutes the functional metalens. The width of each nanopillar varies across the aperture to impart the target phase of each design, providing full 2π phase coverage with high transmission at 532 nm shown in supplementary S3.

### *Fabrication of TUT*

The TUT is fabricated by adapting and optimizing previously established methods (35, 38) to ensure peak acoustic and optical performance. A 250 μm thick, 36° Y-cut lithium niobate ($LiNbO_3$) wafer (Precision Micro-Optics Inc., MA, USA) is selected as the active piezoelectric substrate due to its high electromechanical coupling and superior optical transparency (~80%) at the operating wavelength of 532 nm. Both surfaces of the wafer are deposited with a 200 nm layer of indium tin oxide (ITO) to function as transparent electrodes. The wafer is then diced into 3 mm x 3 mm elements (or 6 mm x 6 mm for large-scale in vivo imaging), after which a thin coaxial wire is bonded to the primary electrode using conductive silver epoxy (8330S, MG Chemicals, Ontario, Canada). The assembly is encased in a slightly oversized brass housing and supported by an acoustic backing layer of transparent epoxy (EPO-TEK 301, Epoxy Technologies Inc., MA, USA). To ensure structural robustness and a planarized surface, a suitably sized cover-glass is adhered to the posterior side of $LiNbO_3$ during the curing process. The brass housing is grounded to the piezoelectric crystal, and a second coaxial wire is attached to the metallic housing with silver epoxy to complete the electrical circuit. The device is then coated with 30 µm Parylene-C (PDS 2010 Labcoter, SCS) as an insulating and matching layer.

***Characterization of TUT***

The optical transparency of TUT is measured ~80% at the operating wavelength of 532 nm using a power meter. The electrical resonance characteristics of the TUT are evaluated using a vector network analyzer (E5100A, Agilent, CA, USA) that measures its response across a range of frequencies. The theoretical center frequency of operation is 14.4 MHz for bare $LiNbO_3$ wafer. However, after incorporating the Parylene-C acoustic matching layer, the experimental response exhibits two distinct resonant peaks at 11.6 MHz and 15.4 MHz, along with corresponding anti-resonant peaks at 11.9 MHz and 15.7 MHz. The observed shift and splitting of the resonance arise from the acoustic loading and impedance modification introduced by the Parylene-C coating. Subsequent acoustic transceiver performance is characterized using a pulse-echo measurement setup. The TUT is then submerged in a chamber filled with deionized water that serves as the acoustic coupling medium. A high voltage pulser/receiver (5073PR, Olympus, Japan) operated at 50 Ω damping, 1 kHz pulse repetition frequency is employed to excite the TUT at its fundamental resonance. The reflected echoes from a polished metal (aluminium) target are recorded using TUT with an impedance matching circuit to calculate the maximum peak to peak voltage of 1.7 V.

***ORPAM experimental setup and data acquisition***

In the experimental setup, a 532 nm pulsed laser (GLPM-10, IPG Photonics) of 1.4 ns pulse duration, 10 kHz repetition rate, 200 nJ pulse energy is employed as the excitation light source. The output laser is sampled using a 10% beam-splitter (BSF10-A, Thorlabs) combined with a photodiode (DET10A, Thorlabs) and it acts as the reference signal. A 16-bit Razormax-16 data acquisition system (Dynamic Signals LLC, Lockport, IL, USA) utilizes the detected beam samples to synchronize the system at a sampling rate of 1 Giga-samples per second. The collimated laser beam illuminates the metalens in a vertical geometry. The phase of the collimated beam is modulated by transmitting through the meta-optical element and creates the desired focused excitation beam tuned by the metalens phase profile. Such configuration features a central aperture allowing for the coaxial transmission of the focused optical beam and simultaneous acoustic detection. It also maximizes the signal-to-noise ratio, sensitivity and effectively minimizes the acoustic loss often associated with off-axis illumination schemes. The optical focus spot formed by metalens is raster scanned with a step size of ~5 µm through the clear TUT surface across the imaging field-of-view using a computer controlled 2-axis translation stage (Physik Instrument). Photoacoustic signals detected by TUT combined with an electrical impedance matching circuit are filtered (3-30 MHz RF bandpass; ZABP-16+, Mini-Circuits) and amplified with two 28-dB amplifiers (ZFL-500LN+, Mini-Circuits) connected in series. At each position the photoacoustic signals are averaged for 5 laser pulses to improve the signal-to-noise ratio. These signals are then digitized and captured by the data acquisition system. The acquisition and image reconstruction scheme is presented in supplementary S7.

### *Phantom preparation*

To prepare the experimental phantoms, a 1.2% (w/v) agar medium is synthesized by dissolving 1.2 g of agar powder in 100 mL of deionized water, chosen for its high optical transparency and acoustic compatibility. The hot agar solution is degassed for 5 minutes to eliminate air-induced bubbles or artifacts. The target leaf is embedded in agar suitably with tweezers whereas carbon fibers (Asbury, MI, USA) and carbon particles (484164, Sigma-Aldrich, USA) are mixed randomly with agar before solidification. The mixture is then transferred to a Petri-dish to initiate cooling. During this phase, the TUT is stabilized using a custom 3D-printed mount (Black Resin V4, Formlabs, USA) and embedded into the agar matrix ensuring the wire electrodes remain accessible. Upon complete solidification of around 10-15 minutes, the electrodes are integrated into the experimental setup for data acquisition.

### *Cranial window preparation for in-vivo imaging*

C57BL/6J mice (Jackson Laboratory Bar Harbor, ME, USA; female, 3-6 months old) are used for the in vivo experiments. All animal procedures are approved by the Institutional Animal Care and Use Committee (IACUC) of The Pennsylvania State University and experiments are performed in accordance with the institutional guidelines. For cranial window surgery, anesthesia is induced using 3% isoflurane mixed with medical oxygen supply (1 L/min) while the animal is secured in a stereotaxic frame. Body temperature is maintained at 37ºC using a warming pad placed below and continuous supply of anesthetic gas (1-3% isoflurane mixed with 1 L/min oxygen) is used during surgery. Hair is removed from the surgical area skin using a depilatory cream, and the exposed skin is disinfected using three alternating cycles of betadine and 70% ethanol. Bupivacaine is injected subcutaneously to provide local anesthesia. A midline scalp incision is then made, and the underlying connective tissue is cleared to expose the skull surface. A rectangular craniotomy window ~$4\times8$ mm$^2$ is created using 0.5 mm dental drill bit (19007-05, Fine Science Tools Inc., California, USA), while regularly cooling the drilled area with sterile saline. The skull flap is gently removed, and the dura is left intact. A thin (80 µm) transparent polymethyl-pentene (PMP) film (MX002, Goodfellow Corp, Utah, USA) is then placed directly over the exposed brain to provide optical clarity, mechanical protection, and long-term stability. The boundary of the cranial window is then sealed with the dental cement. A lightweight metallic (titanium) headbar is also fixed during the same time for awake mice imaging setup. Following the completion of the seal and headbar fixation, the administration of anesthesia is ceased. A removable silicone elastomer (Kwik-Cast, Silicone Bioadhesives, WPI) is applied on the window to protect it from dust or any impurities. The mouse is then moved to a temperature-controlled recovery cage and monitored closely until they regain full consciousness and ambulatory thermoregulation. Following surgery, it is allowed to fully recover for 2 weeks before using for awake-state imaging. During ORPAM imaging, the protective silicone layer was removed, and the animal was head-fixed while awake.

**Funding:** This work was supported by the Penn State Cancer Institute, the NSF CAREER Award EPMD2238878 (S.R.K.), the NIH Cross-Disciplinary Neural Engineering Training Program T32NS115667, and the Washington Research Foundation. Part of this work was conducted at the University of Washington's Washington Nanofabrication Facility / Molecular Analysis Facility, an NNCI site partially supported by the National Science Foundation (awards NNCI-1542101 and NNCI-2025489).

**Author contributions:** S.R.K., A.M., K.S., and J.E.F. contributed to conceptualization. K.S., M.L., J.E.F., S.M., A.M., and S.R.K. developed the methodology. K.S., M.L., S.M., Z.Z., and J.E.F. carried out the investigation. K.S., M.L., S.M., and S.P. performed visualization. A.M., J.E.F., and S.R.K. provided supervision. K.S. wrote the original draft. All authors contributed to review and editing of the manuscript and have given approval to the final version of the manuscript.

**Competing interests:** The authors declare that they have no competing interests.

# Supplementary Information for

## Metalens-transparent ultrasound transducer module for extended depth photoacoustic microscopy

Krishnendu Samanta[1], Myunghoo Lee[2], Shubham Mirg[1,3], Zhihao Zhou[2], Saniya Patil[1], Johannes E. Fröch[2], Arka Majumdar[2,4], and Sri-Rajasekhar Kothapalli[1,3,5]*

[1]Department of Biomedical Engineering, Penn State University, University Park, PA 16802, United States

[2]Department of Electrical Engineering, University of Washington; Seattle, WA 98189, United States

[3]Center for Neural Engineering, Penn State University; University Park, PA 16802, United States.

[4]Department of Physics, University of Washington, Seattle, Washington 98195, United States

[5]Penn State Cancer Institute, Penn State University; Hershey, PA 17033, United States.

*Corresponding author. Email: srkothapalli@psu.edu

**This PDF file includes:**

Figs. S1 to S12

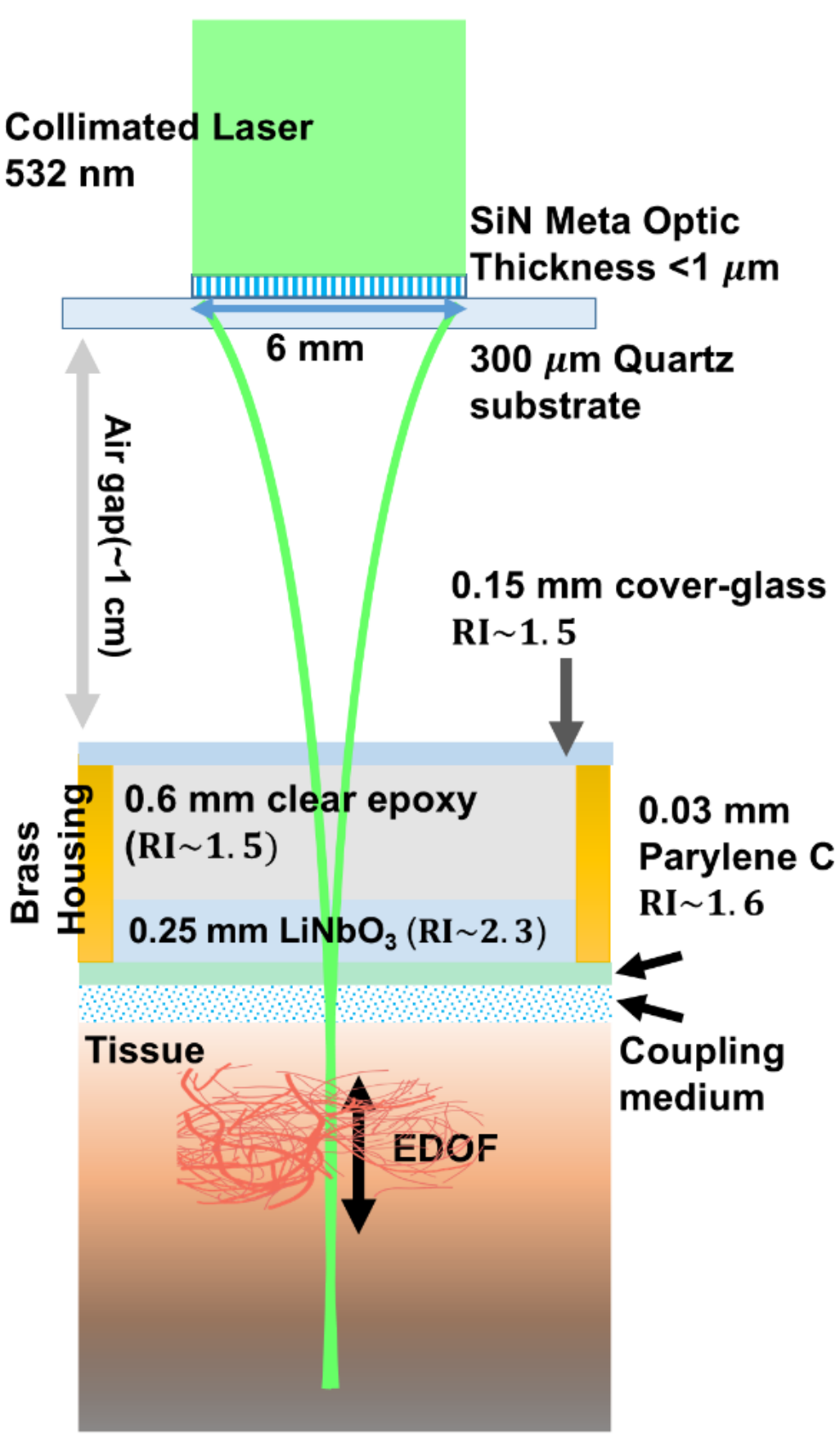

*Fig. S1. Optical path through the metalens-TUT setup: Cross sectional schematic showing transmission of focused excitation beam through the individual layers of TUT and acoustic coupling medium to the imaging target. TUT: transparent ultrasound transducer. RI: Refractive Index.*

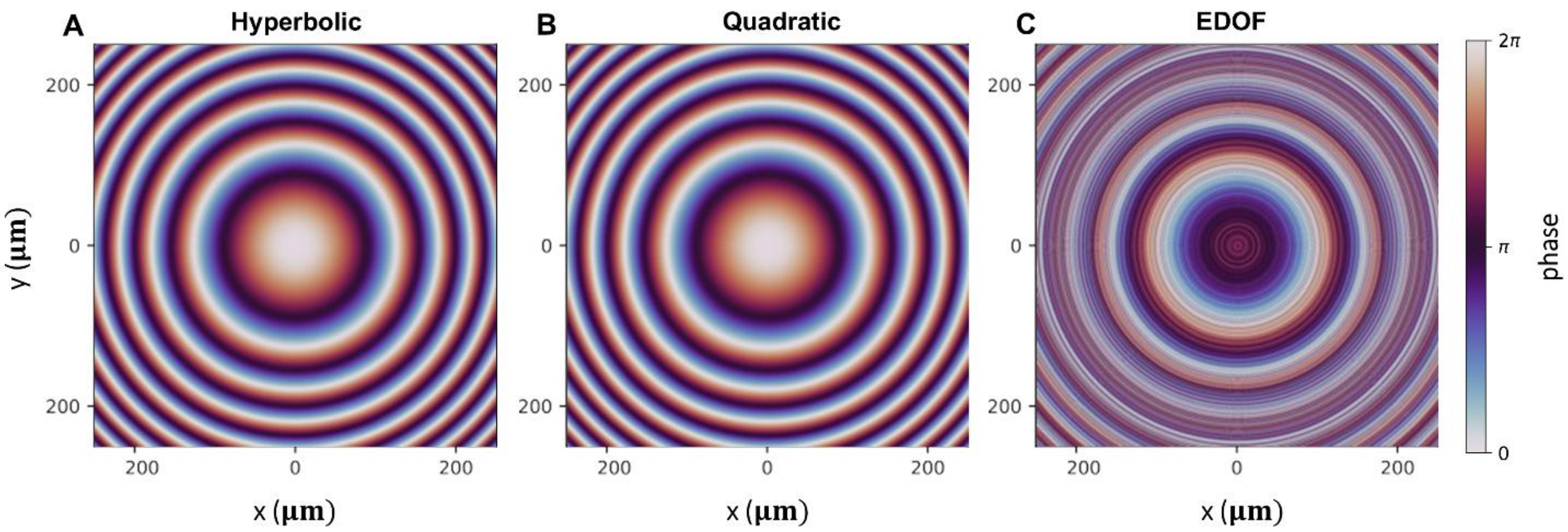

*Fig. S2. Wrapped phase profiles of the three metalens designs: Wrapped phase maps of the central region (Ø0.5 mm) of the (A) hyperbolic, (B) quadratic, and (C) EDOF metalenses. The hyperbolic and quadratic-phase designs exhibit regular concentric phase variations, whereas the optimized EDOF metalens exhibits an aperiodic phase distributions designed to generate an extended axial focus.*

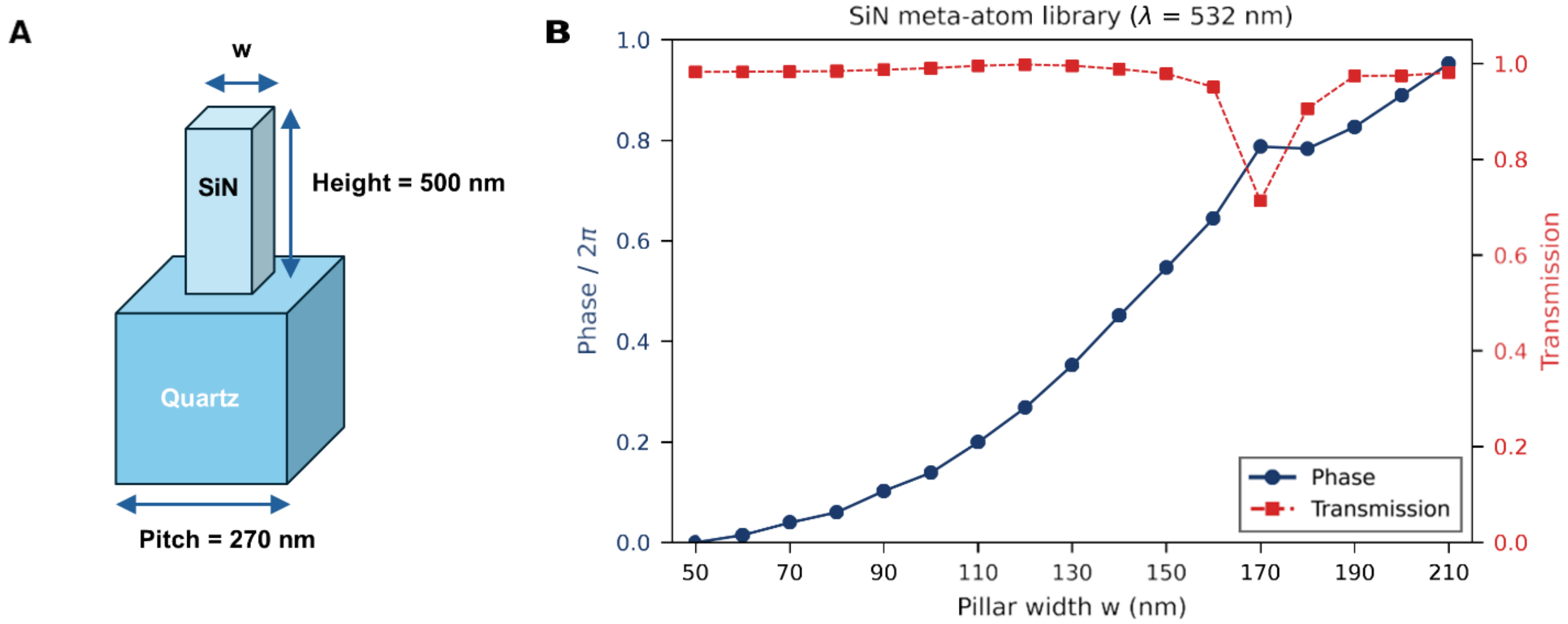

**Fig. S3. Meta-atom library used for metalens designs:** (**A**) Schematic of the square silicon nitride (SiN) nanopillar unit cell on a quartz substrate with pitch = 270 nm, height = 500 nm, and variable width w. (**B**) Simulated phase response and optical transmission as functions of nanopillar width at $\lambda = 532$ nm. Pillar width ranging from 50 nm to 210 nm are sampled in 10-nm steps, providing full $2\pi$ phase coverage while maintaining high optical transmission.

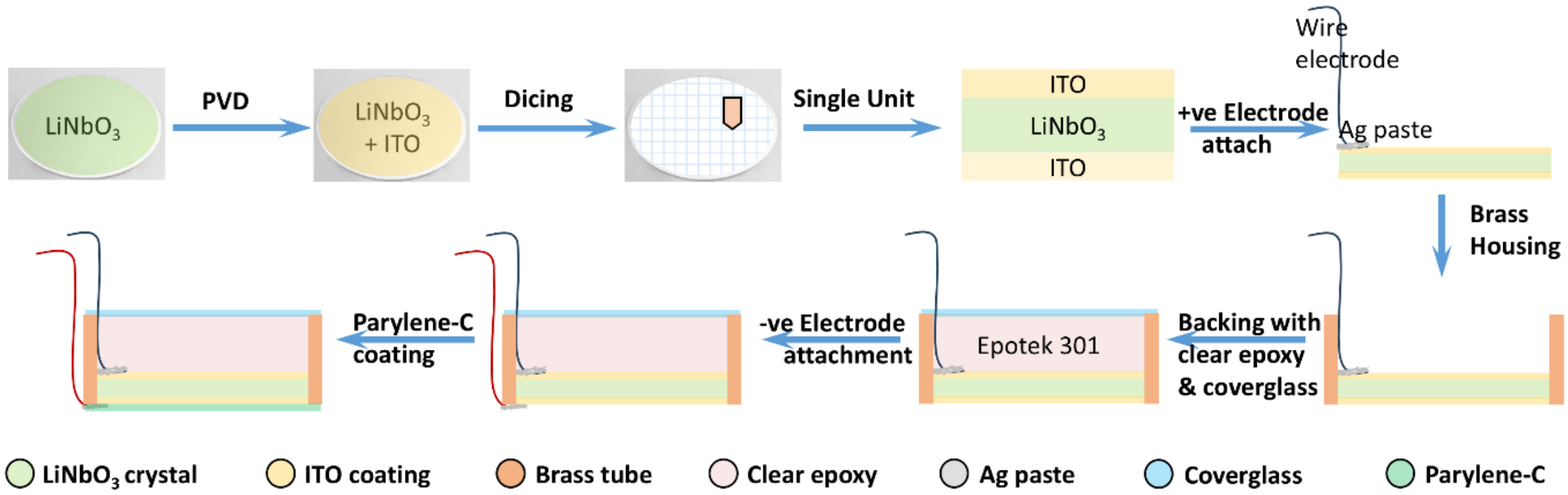

**Fig. S4. Fabrication workflow of transparent ultrasound transducer (TUT):** A 250 µm thick, 36° Y-cut $LiNbO_3$ wafer is coated on both surfaces with 200 nm transparent indium tin oxide (ITO) via physical vapor deposition (PVD) and diced into 3 mm x 3 mm (or 6 mm x 6 mm) elements. A coaxial wire is electrically connected to one ITO electrode surface using conductive silver epoxy and the $LiNbO_3$ is encased in a brass housing. A transparent epoxy backing layer and posterior cover glass provides mechanical support. The opposite electrode and brass housing are electrically connected to complete the transducer assembly, followed by the deposition of a thin Parylene-C layer for electrical insulation and acoustic impedance matching.

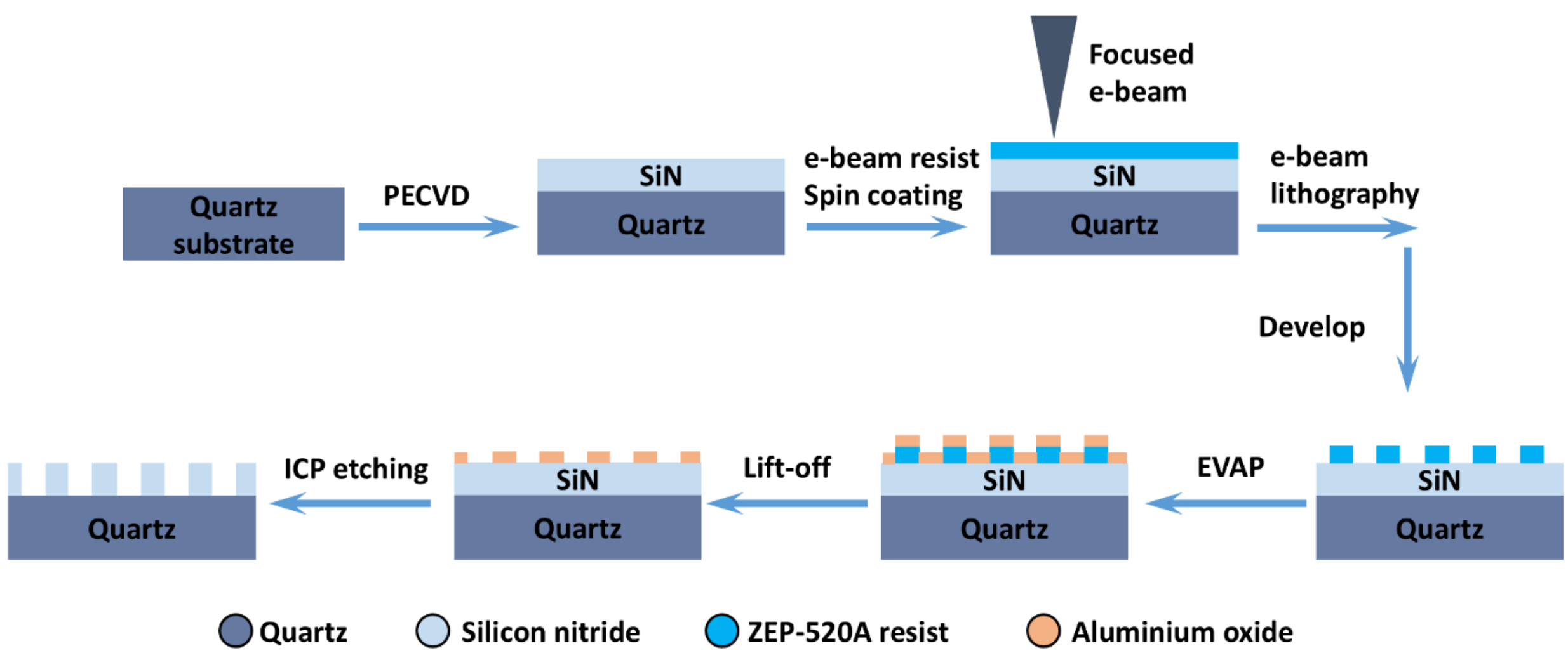

**Fig. S5. Fabrication workflow of the silicon nitride metalens:** A silicon nitride (SiN) film is deposited on a quartz substrate via plasma enhanced chemical vapor deposition (PECVD), followed by spin coating with ZEP-520A electron-beam resist. The metalens pattern is defined by electron beam lithography (EBL) and developed, after which an aluminium oxide hard mask is deposited by electron beam evaporation (EVAP) and defined via lift-off. The nanostructure pattern is subsequently transferred into the SiN layer by inductively coupled plasma (ICP) etching to form the nanopillar array.

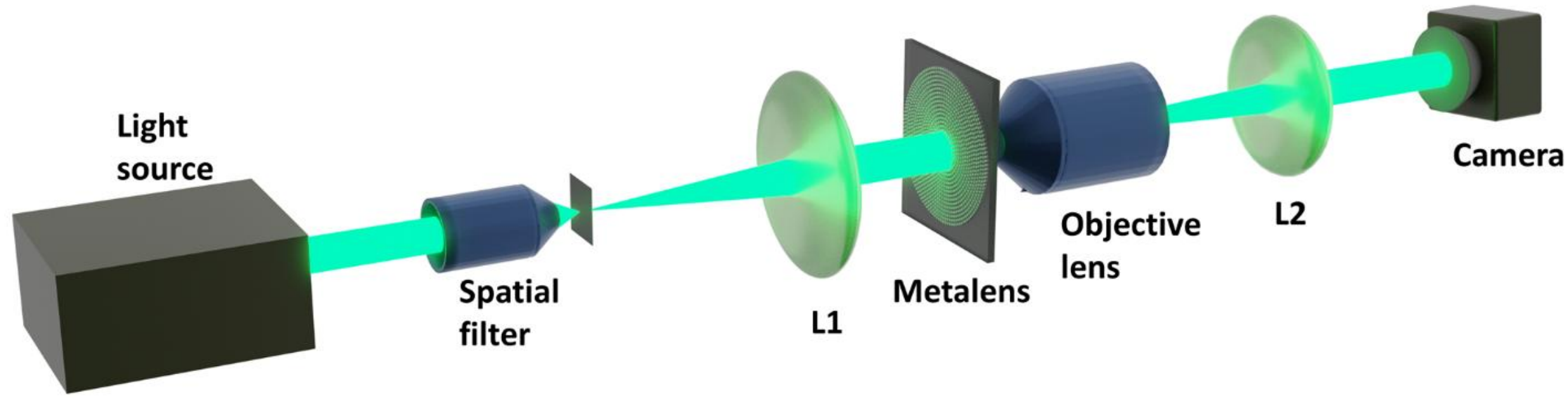

**Fig. S6. Experimental setup for optical characterization of the metalenses:** A spatially filtered collimated beam (532 nm) is incident on the metalens and focused according to its phase profile. The resulting focal optical spot is characterized using a custom microscope comprising an objective lens, tube lens, and camera. Lateral intensity distributions are recorded at successive axial positions by translating the system along propagation direction to estimate the focal width and depth-of-focus.

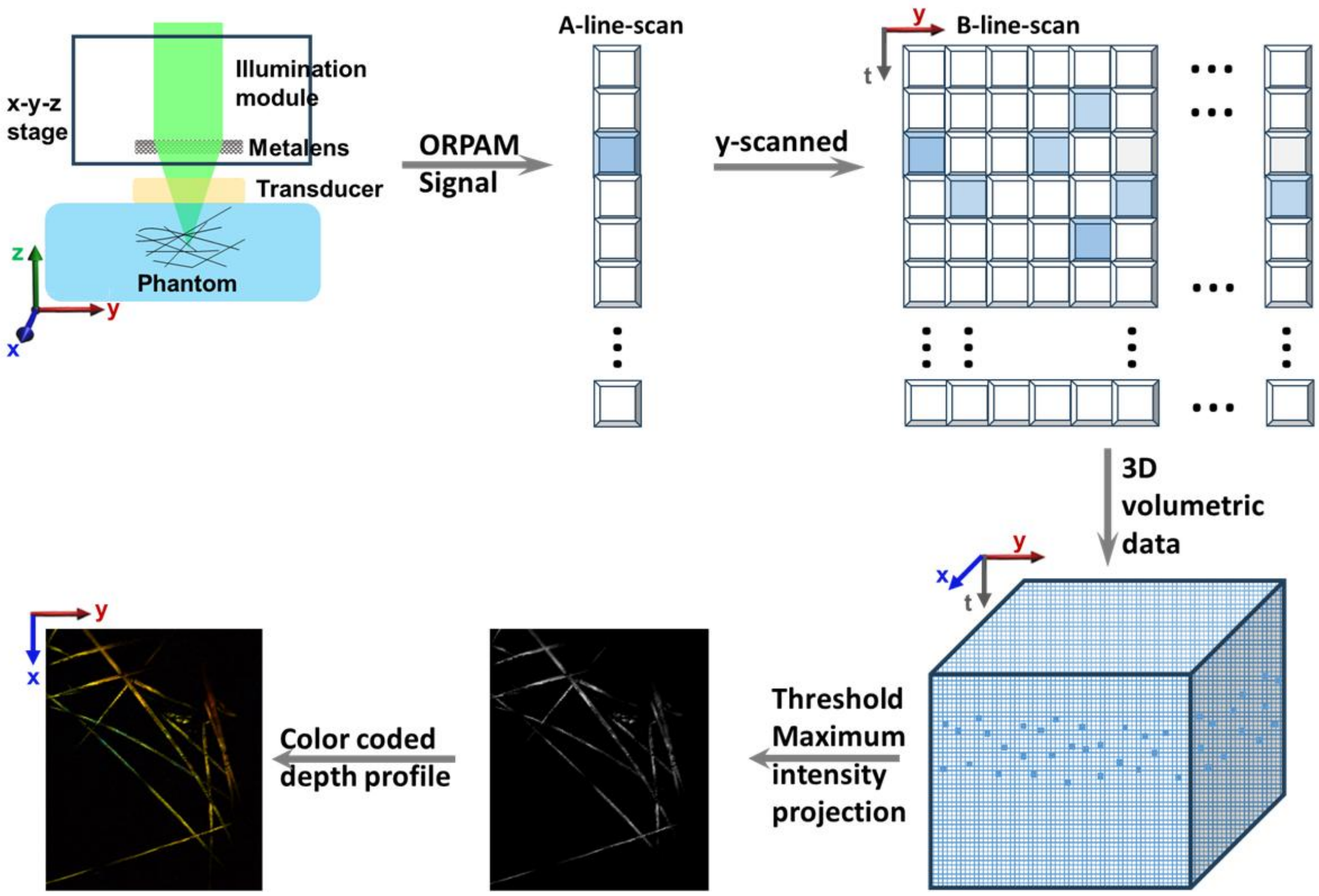

**Fig. S7. ORPAM data acquisition and image reconstruction workflow:** Pulsed optical excitation through the metalens-TUT architecture generates photoacoustic signals from the sample which is detected by the TUT to produce an A-line at each lateral position. One-dimensional lateral scanning assembles successive A-lines into a cross-sectional B scan, whereas two-dimensional raster scanning in x–y plane generates a three-dimensional photoacoustic dataset. Maximum-intensity-projection images are depth-encoded subsequently from the volumetric dataset.

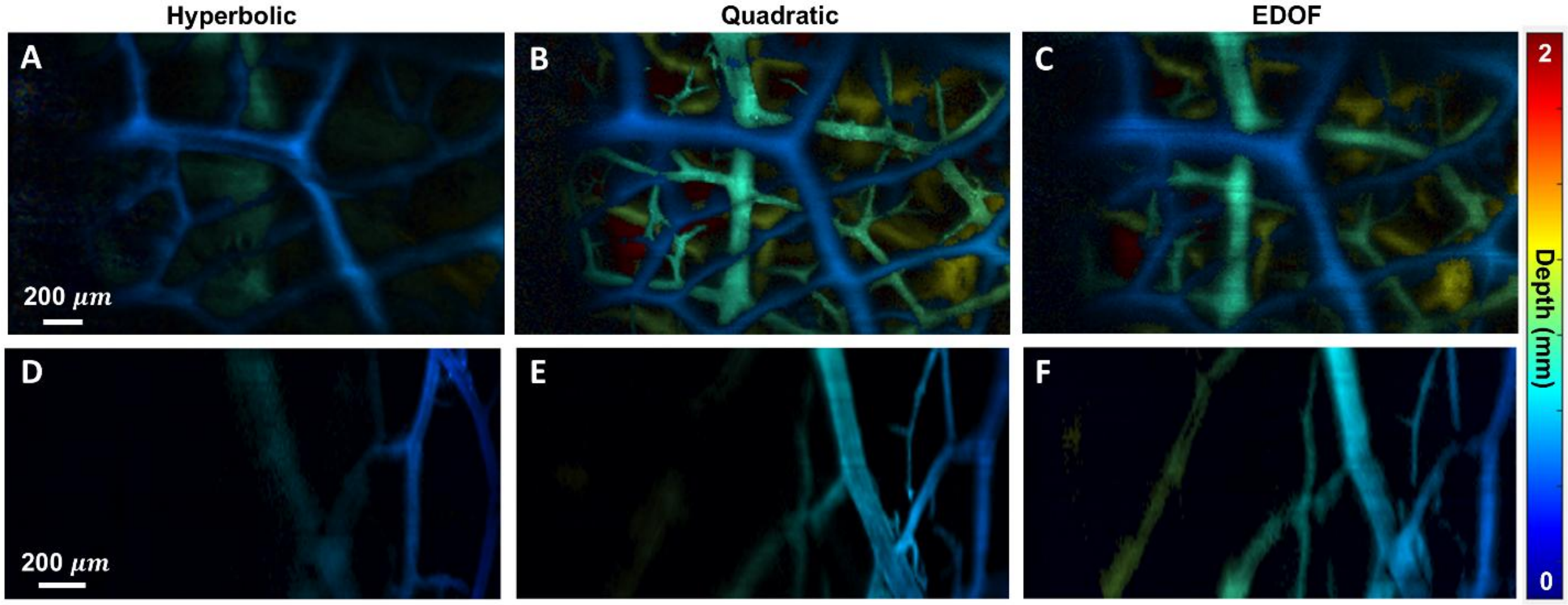

**Fig. S8. Additional ORPAM imaging of multi-layer planar and single-layer inclined leaf phantoms:** (**A** to **C**) Depth-encoded ORPAM images of a multi-layer planar leaf embedded in agar acquired using (**A**) hyperbolic, (**B**) quadratic, and (**C**) EDOF metalenses. (**D** to **F**) depth-encoded photoacoustic images of a single-layer inclined leaf in agar using (**D**) hyperbolic, (**E**) quadratic, and (**F**) EDOF metalenses.

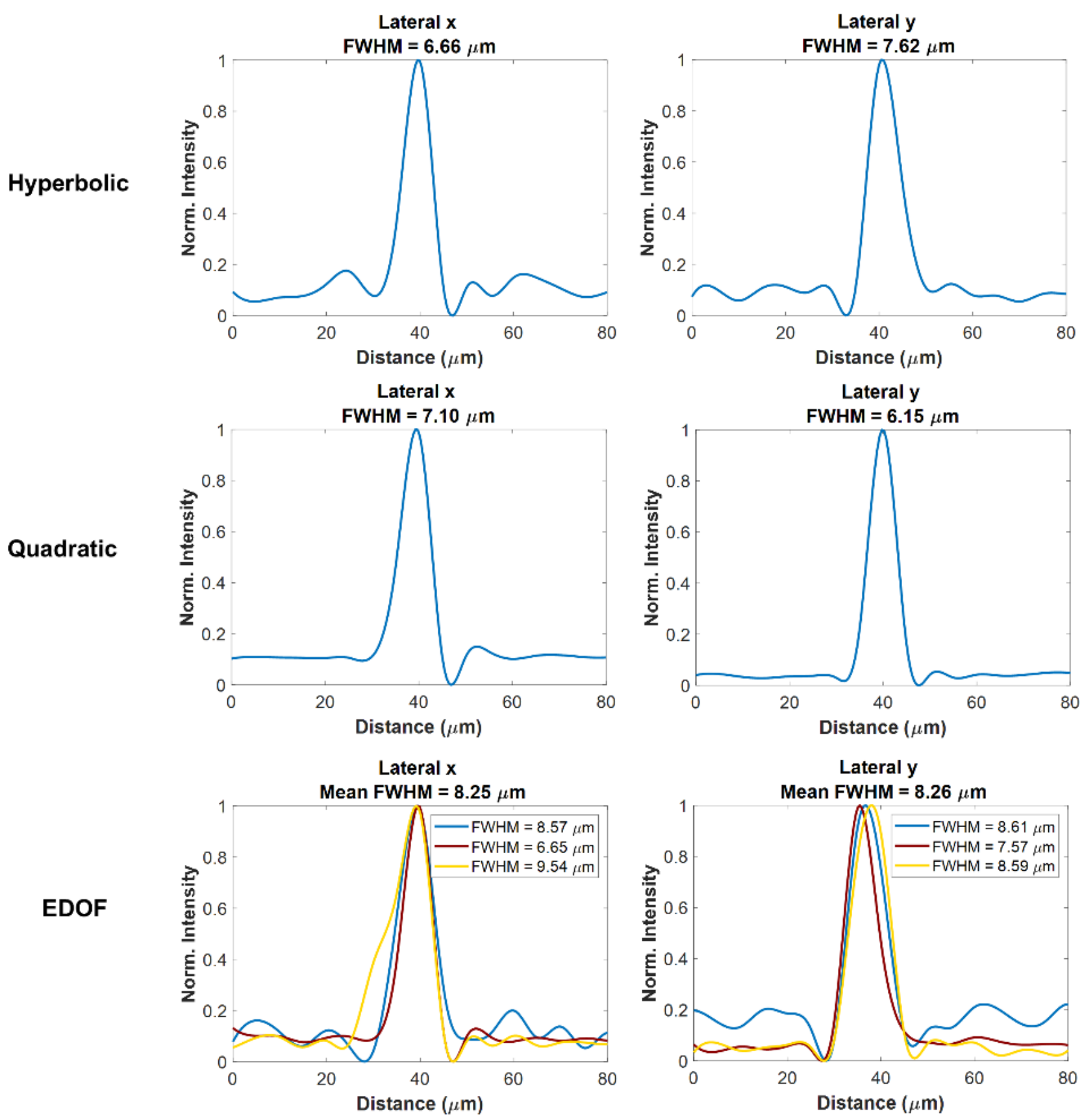

**Fig. S9. Lateral signal-width characterization from carbon micro-particles:** Representative lateral x and y intensity profiles with corresponding full width at half maxima (FWHM) from carbon micro-particle signals acquired using hyperbolic, quadratic, and EDOF metalens configurations. Measurements for hyperbolic and quadratic metalenses are performed at the center of the focal depth, whereas measurements for EDOF are evaluated at the center and two extreme focal positions to demonstrate extended depth-of-focus performance.

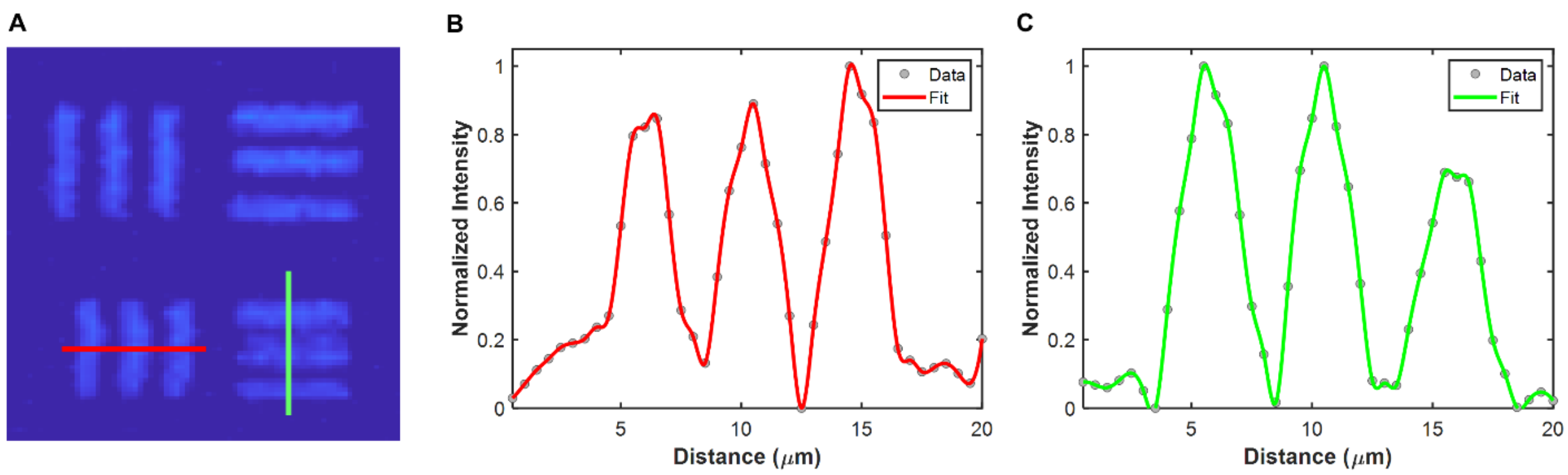

**Fig. S10. Line profile analysis of USAF resolution-target imaging:** (**A**) Selected region of the ORPAM image of a USAF resolution target; (**B**) horizontal and (**C**) vertical line profiles extracted across Group 7, Element 6 of the USAF target ORPAM image obtained using the EDOF metalens.

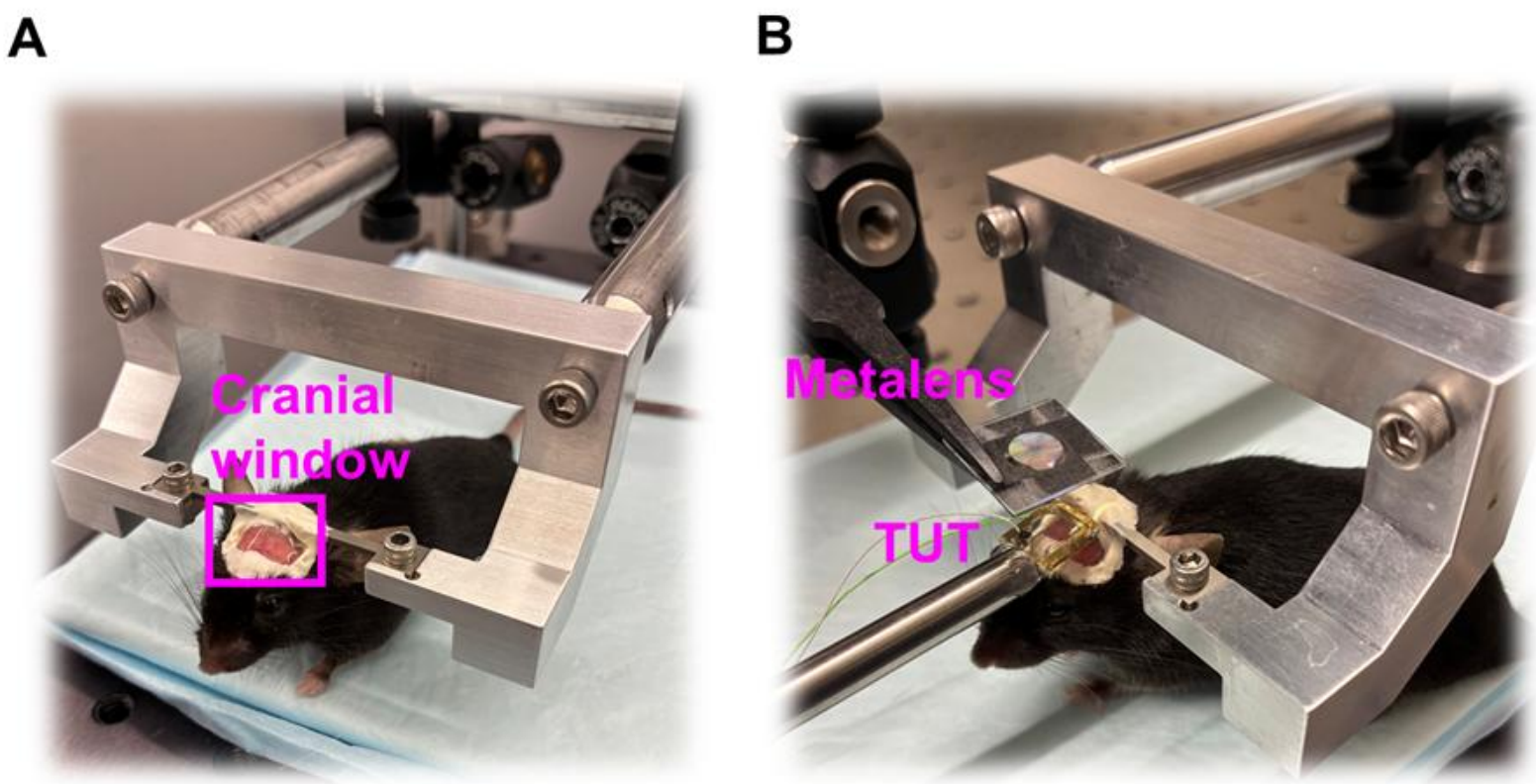

**Fig. S11. Experimental preparation for ORPAM imaging for an awake, head-fixed mouse:** (**A**) Awake mouse secured head-fixation using the implanted titanium headbar, showing exposed cranial window after removal of the protective silicone cover; (**B**) placement of transparent ultrasound transducer (TUT) on cranial window using 1.2% agar for acoustic coupling medium. The metalens excitation path is coaxially aligned with the TUT detection path for ORPAM imaging of cortical vasculature structures.

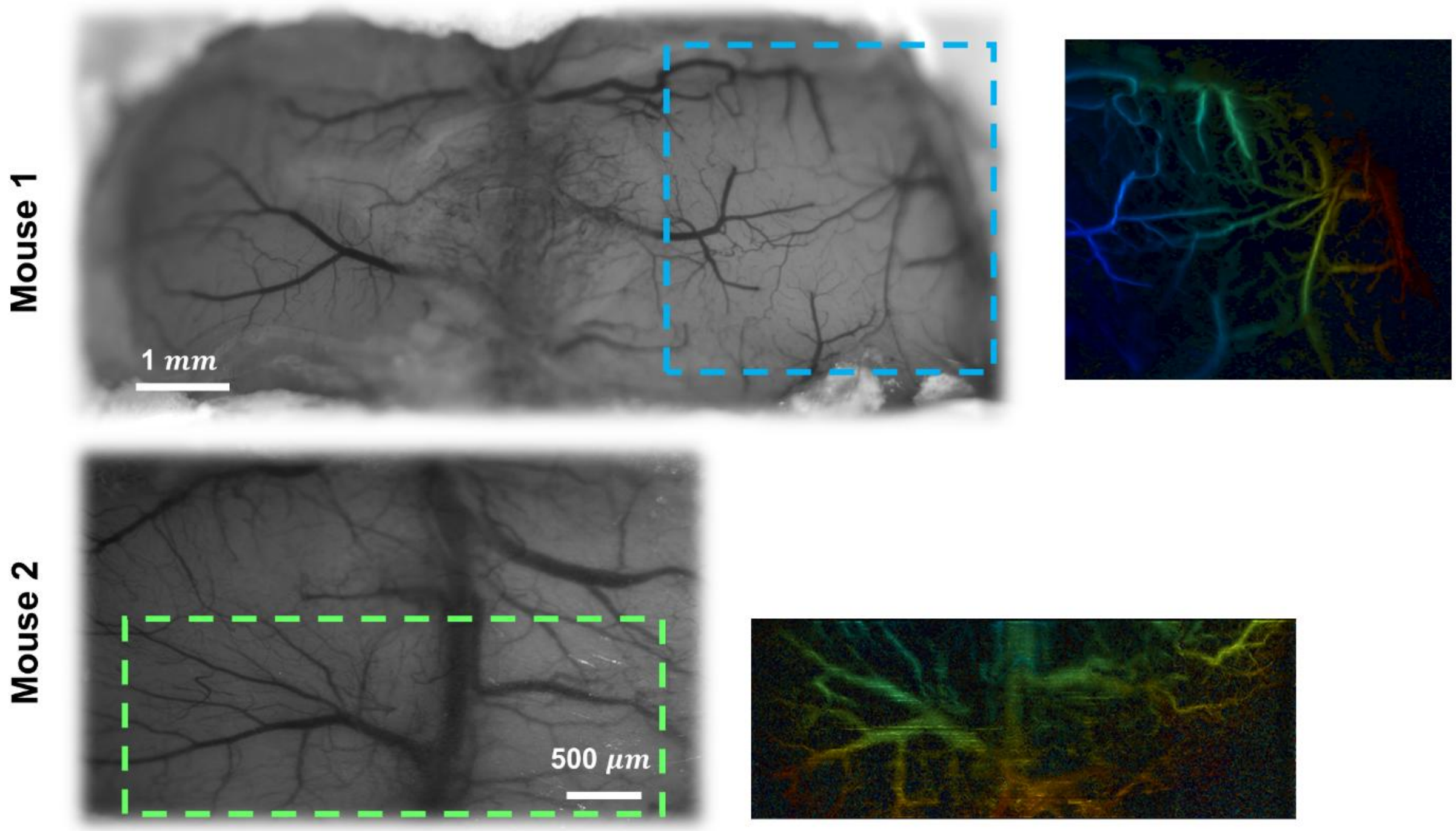

**Fig. S12. Brightfield and ORPAM imaging of cerebral vasculature structure in additional mice:** Brightfield optical images of cranial window regions and corresponding depth-encoded ORPAM images acquired using the EDOF metalens for two different mice, dashed boxes indicate the regions of interest imaged by ORPAM.